\documentclass[useAMS,usenatbib]{mn2e}
\usepackage{epsfig}
\usepackage{graphicx,rotating,amssymb,supertabular,longtable,lscape}
\usepackage{amsmath}
\usepackage{textcomp}

\def\mum{\textmu m}
\def\mums{\textmu m }

\title[Dust tori in AGN]
{Properties of dusty tori in active galactic nuclei - II. Type 2 AGN}

\author[Hatziminaoglou, Fritz \& Jarrett]{E. Hatziminaoglou$^{1}$, J. Fritz$^{2}$,
T.H. Jarrett$^{3}$\\
$^{1}$European Southern Observatory, Karl-Schwarzschild-Str. 2, 85748 Garching bei M\"unchen, Germany\\
$^{2}$INAF/Astronomical Observatory of Padova, Vicolo dell'Osservatorio 2, 35122 Padova, Italy\\
$^{3}$IPAC, California Institute of Technology, 770 South Wilson Avenue, Pasadena, CA 91125, USA}
\begin{document}

\maketitle

\begin{abstract}
This paper is the second part of a work investigating the properties of
dusty tori in active galactic nuclei (AGN) by means of multi-component 
spectral energy distribution (SED) fitting. 
It focuses on low luminosity, low redshift ($z \le 0.25$)
AGN selected among emission line galaxies in the overlapping regions between SWIRE and SDSS Data 
Release 4 as well as X-ray, radio and mid-infrared selected type 2 AGN samples from the literature.
The available multi-band photometry covers the spectral range
from the $u$-band up to 160\mum. Using a standard $\chi^2$ minimisation, the observed SED
of each object is fit to a set of multi-component models comprising
a stellar component, a high optical depth ($\tau_{9.7} \ge 1.0$) torus and cold emission from a starburst (SB).
The torus components assigned to the majority of the objects were those of the highest
optical depth of our grid of models ($\tau_{9.7}=10.0$).
The contribution of the various components (stars, torus, SB) is reflected
in the position of the objects on the IRAC colour diagram, with star-, torus- and starburst-dominated
objects occupying specific areas of the diagrams and composite objects lying in between.
The comparison of type 1 (as derived from Paper 1, \citealt{hatzimi08}) 
and type 2 AGN properties is broadly consistent with the Unified Scheme.
The estimated ratio between type 2 and type 1 objects is about 2-2.5:1.
The AGN accretion-to-infrared luminosity ratio is an indicator of the obscuration of the AGN
since it scales down with the covering factor. 
We find evidence supporting the receding torus paradigm, with the
estimated fraction of obscured AGN, derived from the distribution of the covering factor, decreasing with increasing optical luminosity ($\lambda L_{5100}$)
over four orders of magnitude. 
The average star formation rates are of $\sim 10 M_{\odot}/yr$
for the low-$z$ sample, $\sim 40 M_{\odot}/yr$ for the other type 2 AGN and $\sim 115 M_{\odot}/yr$
for the quasars; this result however, might simply reflect observational biases, as the quasars
under study were one to two orders of magnitude more luminous than the various type 2 AGN.
For the large majority of objects with 70 and/or 160 \mums detections an SB
component was needed in order to reproduce the data points, implying that the far-infrared 
emission in AGN arises mostly from star formation; moreover,
the starburst-to-AGN luminosity ratio shows a slight trend with increasing luminosity.

\end{abstract}


\section{INTRODUCTION}
\label{intro}

Dust is the cornerstone of the active galactic nuclei (AGN) 
Unified Scheme that postulates that the
diversity of the observed properties of AGN are merely a result of the
different lines of sight with respect to obscuring material surrounding the
active nucleus (\citealt{antonucci93}; \citealt{urri95}; \citealt{tadhunter08}). 
The elements necessary for understanding the nature of AGN
and the variety of their properties are in fact the geometry of the 
circumnuclear dust and the amount of obscuration. The latter is also needed in order to understand the 
physical conditions of the dust enshrouded
nuclear region and to estimate the intrinsic UV-to-optical properties of the AGN.
Obscuring dusty material should re-radiate in the infrared (IR) the fraction of accretion
luminosity it absorbs, providing thus a direct metric for the study of the medium.
Indeed all spectroscopically confirmed AGN lying in regions covered by the various Spitzer
Space Telescope (Spitzer) surveys show significant IR emission down to the detection limits.

The conventional wisdom is that dust, distributed in a more or less toroidal form 
around the nucleus, mainly consists of silicate and graphite grains,
leaving unmistakable signatures in the observed SEDs of AGN. 
The most important dust features are the T$\simeq$1500 black-body like
rise of the IR emission at $\lambda \ge 1$ \mum, corresponding to the sublimation
temperature of the graphite grains and the
$\sim$9.7 micron feature in emission/absorption attributed to silicates.
While silicates in absorption were already long ago observed in a variety of type 2
objects, ISO observations only provided faint indications of their presence in 
emission in the spectra of type 1, challenging the Unified Scheme, as all smooth
torus models predicted a feature in emission for the unobscured AGN.
Clumpy models, on the other hand, provided a natural attenuation of the feature
\citep{nenkova02}. Observations of tens of objects carried out with the
IRS spectrograph onboard Spitzer allowed, however,
the detection of the silicate feature (first reported by \citealt{siebenmorgen05}
and observed by several others since)
predicted by smooth tori models in emission for type 1 AGN and quasars
(e.g. \citealt{pier92}; \citealt{granato94}; \citealt{fritz06}).
Thus far, there is no consensus on the dust distribution among astronomers;
for an overview of the issues involving the two approaches see \cite{dullemond05},
\cite{elitzur08} and \cite{nenkova08}.

Until very recently the number of spectroscopically confirmed type 2 AGN
was rather limited; in the last few years however, the advent of large-scale multi-wavelength 
surveys has led to a revolution in AGN studies. We are now in a position to
compile statistically large type 2 AGN samples at all redshifts, selected in a multitude of ways.
In this work we try to characterise the observed spectral energy distribution (SED) of
type 2 AGN using the smooth torus models described in \cite{fritz06} and
the methodology described in \cite{hatzimi08}, hereafter Paper 1, in which we dealt with the
IR properties of quasars. The aim of the present work is to model a large sample of type 2 AGN and, by comparison
with the results derived from the study of type 1 quasars in Paper 1, test the Unified Scheme.
The various type 2 AGN samples will be discussed in detail in Section
\ref{sec:data}. Section \ref{sec:seds} will briefly describe the various model components
as well as the basics of the standard $\chi^2$ minimisation technique and assumptions.
The results on the individual samples will be presented in Section \ref{sec:fitresults} while
the combined results from all samples, including the quasars (typical representatives of
type 1 AGN) from Paper 1, will be shown
in Section \ref{sec:combi}. A final discussion on the future of such studies is presented
in Section \ref{sec:discuss}.

\section{THE SAMPLES}
\label{sec:data}

In this work we will be analysing a variety of type 2 AGN samples, 
namely a low redshift, low luminosity AGN sample, an X-ray and radio
selected sample in the COSMOS field and a mid-infrared (MIR) selected sample
from the ELAIS fields. We will be also revisiting the SDSS/SWIRE quasar
sample studied in Paper 1 (hereafter quasar sample).

\subsection{Low-luminosity low-redshift AGN}

The low-luminosity low-redshift AGN sample, hereafter low-$z$ sample,
is part of an updated version of an emission line galaxies sample 
classified as AGN. The selection was made among hundreds of thousands of emission
line galaxies from the SDSS Data Release 4 (DR4; \citealt{adelman06}),
based on the relative strength of the emission lines on a
log({\sc [Oiii]}/H$\beta$) - log({\sc [Nii]}/H$\alpha$) diagram, after
careful subtraction of the stellar continuum, described in detail on
\cite{kauffmann03}
\footnote{http://www.mpa-garching.mpg.de/SDSS/DR4/Data/agncatalogue.html}.
From the more than 80,000 proposed AGN, 420 reside in the
northern SWIRE (\citealt{lonsdale03}; \citealt{lonsdale04})
fields ELAIS N1, ELAIS N2 and Lockman. Among these objects,
388 have been detected in at least two out of the four IRAC bands, 
with 340 having also MIPS detections at 24 \mum
\footnote{The four IRAC and three MIPS channels' effective wavelengths are
3.6, 4.5, 5.8 and 8.0 \mums and 24, 70 and 160 \mum, respectively}.
From the 388 (340 with an additional 24 \mums detection) objects,
155 (146) and 88 (81) were additionally detected by MIPS at 70 and 160 \mum, 
respectively. The far-infrared (FIR)
measurements characterize or constrain (with upper limits) the cold dust emission.

\begin{figure}
\centerline{
\psfig{file=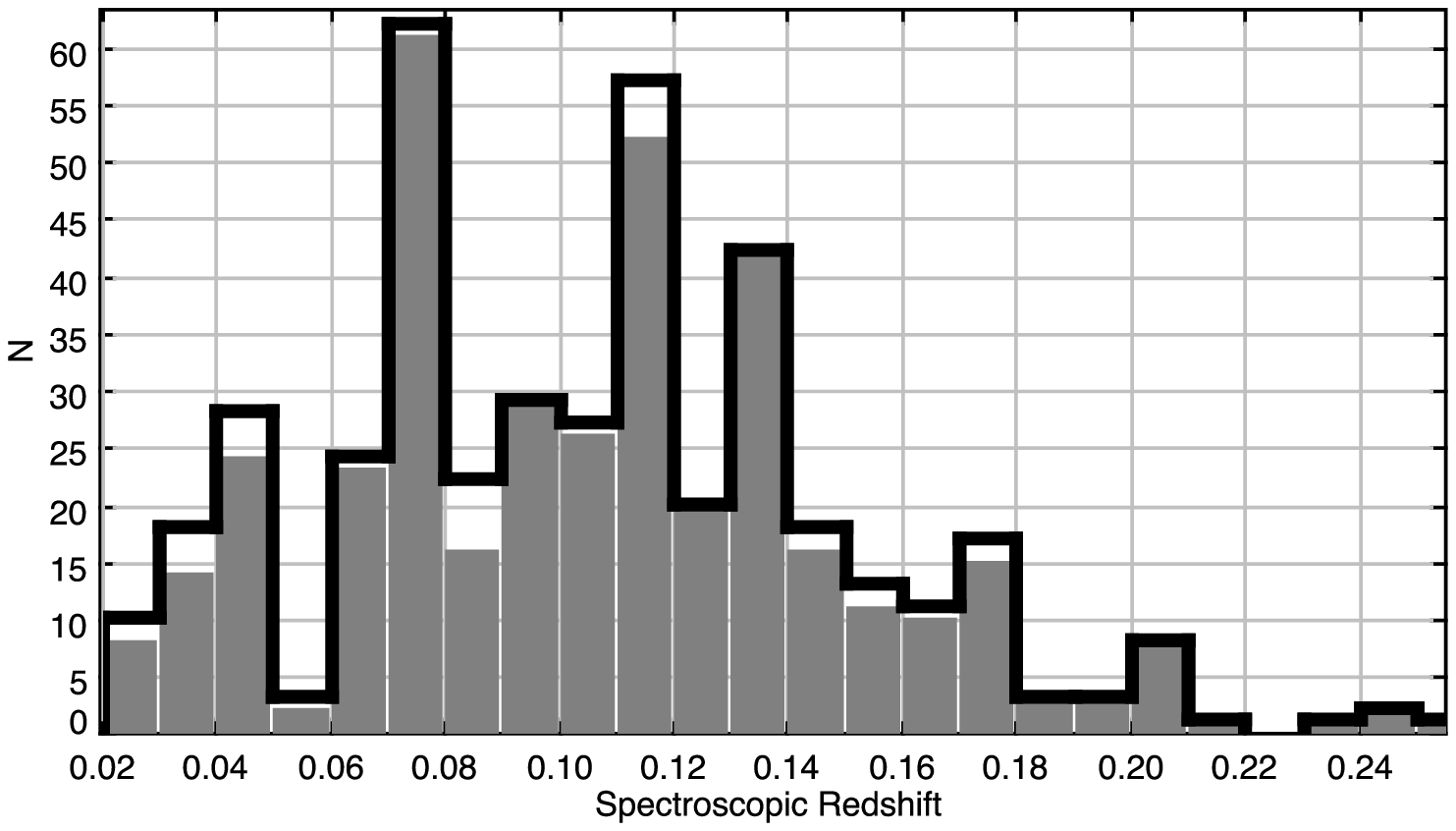,width=10cm}}
\caption{Redshift histogram of the low luminosity low redshift AGN sample
selected from Kauffmann et al. 2003.  The grey shading indicates sources
with good photometric coverage allowing detailed study.}
\label{fig:zhisto1}
\end{figure}

Due to the nature of the sample, particular care was taken in the construction
of the SEDs. We used data from the SDSS DR4, 2MASS including
the deeper data set 2MASS $\times$ 6 taken in the Lockman field \citep{beichman03},
as well as IR data from SWIRE. For extended objects (based on the SDSS
classification)
we used Petrosian $u$, $g$, $r$, $i$, $z$ magnitudes, 7" radius magnitudes for 2MASS $JHK$,
and aperture 5 fluxes (5.8" radius), the largest publicly available IRAC aperture, 
for the four IRAC bands, to better capture
the total light arising from the resolved host galaxies. 
The 5" 2MASS magnitudes, closer to the 5.8" aperture selected for the IRAC bands,
were proven to be inadequate when all the SEDs were visually inspected
as they were clearly falling short of both the SDSS and the IRAC fluxes.
For the point sources we used PSF SDSS magnitudes, default (PSF) 2MASS magnitudes
and aperture 2 (1.9") IRAC magnitudes (for a full description of the SWIRE magnitudes
and their adequate for point-like and extended sources see \citealt{surace05}).
The redshift histogram of the 420 objects is shown in Fig. \ref{fig:zhisto1}, 
with the gray histogram corresponding to the 388 objects that will be studied in detail.
Tables \ref{tab:sdssObj} and \ref{tab:swireFluxes} show the SDSS names, coordinates, 
redshifts, emission line properties, and stellar masses (taken from the narrow
emission line object catalogue), and SWIRE fluxes, respectively
for the 420 objects. The objects belonging to the Subset (S) of 388 entries used 
for the study are marked in the column
``S'' of Table \ref{tab:sdssObj}. The objects marked in column ``AGN'' are
those that were assigned with an AGN component as a result of the fitting procedure.

\begin{table*}
\caption{SDSS names, coordinates, redshifts, emission line properties, and stellar
masses for the 420 objects overlapping the SWIRE fields and the narrow emission line
object catalogue from Kauffmann et al (2003). Column called ``S" (for Subset) marks the objects
that have an adequate sampling of their SEDs and will be used in the analysis.
The full catalogue is available online.}
\label{tab:sdssObj}
\begin{tabular}{|r|l|l|r|r|r|r|r|r|r|}
\hline
  \multicolumn{1}{|c|}{SqNr} &
  \multicolumn{1}{c|}{S} &
  \multicolumn{1}{c|}{SDSS name} &
  \multicolumn{1}{c|}{RA} &
  \multicolumn{1}{c|}{Dec} &
  \multicolumn{1}{c|}{$z$} &
  \multicolumn{1}{c|}{log[OIII]} &
  \multicolumn{1}{c|}{log[OIII/Hb]} &
  \multicolumn{1}{c|}{log[NII/Ha]} &
  \multicolumn{1}{c|}{logM$^*$[M$_{\odot}$]}\\
\hline
 1 &       & SDSS\_J160215.17+535456.1 & 16:02:15.17 & +53:54:56.1 & 0.081 & 6.2711 & 0.341 & -0.0537 & 11.065     \\
  2 &   X  & SDSS\_J160258.45+540929.1 & 16:02:58.45 & +54:09:29.1 & 0.065 & 5.6622 & 0.113 & 0.1397 & 10.4765     \\
  3 &   X  & SDSS\_J160300.04+541737.3 & 16:03:00.04 & +54:17:37.3 & 0.106 & 5.862 & 0.0946 & -0.0224 & 11.1844    \\
  4 &   X  & SDSS\_J160221.35+541105.0 & 16:02:21.35 & +54:11:05.0 & 0.066 & 6.4168 & -0.0782 & -0.2444 & 11.2855  \\
  5 &       & SDSS\_J160006.83+541717.2 & 16:00:06.83 & +54:17:17.2 & 0.066 & 6.4527 & -0.115 & -0.2069 & 10.5041  \\
  6 &   X  & SDSS\_J160114.70+542034.3 & 16:01:14.70 & +54:20:34.3 & 0.064 & 7.7295 & -0.1227 & -0.2971 & 10.7284  \\
  7 &   X  & SDSS\_J160035.48+541529.2 & 16:00:35.48 & +54:15:29.2 & 0.064 & 6.1645 & 0.4281 & -0.1508 & 10.7402   \\
  8 &       & SDSS\_J155713.43+543953.9 & 15:57:13.43 & +54:39:53.9 & 0.045 & 5.9737 & -0.0429 & -0.0453 & 10.712  \\
  9 &   X  & SDSS\_J155813.02+543656.1 & 15:58:13.02 & +54:36:56.1 & 0.047 & 5.6459 & -0.2685 & -0.2906 & 10.4376  \\
  10 &       & SDSS\_J155705.98+543902.0 & 15:57:05.98 & +54:39:02.0 & 0.045 & 6.261 & -0.0665 & 0.0228 & 10.8279  \\
  11 &       & SDSS\_J155708.97+544355.7 & 15:57:08.97 & +54:43:55.7 & 0.047 & 5.6832 & 0.1481 & 0.0986 & 10.7445  \\
  12 &   X  & SDSS\_J160007.28+545350.0 & 16:00:07.28 & +54:53:50.0 & 0.084 & 7.264 & -0.3575 & -0.247 & 10.5545   \\
  13 &   X  & SDSS\_J155940.66+545050.8 & 15:59:40.66 & +54:50:50.8 & 0.093 & 6.031 & 0.2662 & 0.0451 & 10.9104    \\
  14 &   X  & SDSS\_J160218.41+544718.5 & 16:02:18.41 & +54:47:18.5 & 0.083 & 6.887 & 0.0743 & -0.1568 & 10.6365   \\
  15 &   X  & SDSS\_J160254.28+545540.7 & 16:02:54.28 & +54:55:40.7 & 0.120 & 6.5346 & 0.3394 & 0.1132 & 11.1944   \\
  16 &   X  & SDSS\_J160124.58+550120.1 & 16:01:24.58 & +55:01:20.1 & 0.037 & 7.1557 & 0.7 & 0.0261 & 10.6611     \\
  17 &   X  & SDSS\_J160512.63+545550.7 & 16:05:12.63 & +54:55:50.7 & 0.064 & 6.0292 & 0.7485 & -0.0783 & 9.9791    \\
  18 &   X  & SDSS\_J160122.30+552213.6 & 16:01:22.30 & +55:22:13.6 & 0.119 & 6.3901 & -0.3305 & -0.1667 & 10.9273  \\
  19 &       & SDSS\_J161320.00+525516.4 & 16:13:20.00 & +52:55:16.4 & 0.166 & 8.0556 & 0.3297 & -0.1401 & 11.2084  \\
  20 &   X  & SDSS\_J161339.10+531833.6 & 16:13:39.10 & +53:18:33.6 & 0.107 & 7.3537 & -0.2625 & -0.2766 & 11.1697  \\
\hline\end{tabular}
\end{table*}

\begin{table*}
\caption{SWIRE fluxes of the sample presented in Table \ref{tab:sdssObj}.
The typical errors on the 70 and 160 \mums fluxes are 10\% of the fluxes.
The column ``AGN" denotes the objects that, after the fitting,
were found to have an AGN component and on which all AGN-related analysis is based.
The full sample is available online.}
\label{tab:swireFluxes}
\begin{tabular}{|r|r|r|r|r|r|r|r|l|}
\hline
  \multicolumn{1}{|c|}{SqNr} &
  \multicolumn{1}{c|}{S$_{3.6}$ [$\mu$Jy]} &
  \multicolumn{1}{c|}{S$_{4.5}$ [$\mu$Jy]} &
  \multicolumn{1}{c|}{S$_{5.8}$ [$\mu$Jy]} &
  \multicolumn{1}{c|}{S$_{8.0}$ [$\mu$Jy]} &
  \multicolumn{1}{c|}{S$_{24}$  [$\mu$Jy]} &
  \multicolumn{1}{c|}{S$_{70}$  [$\mu$Jy]} &
  \multicolumn{1}{c|}{S$_{160}$ [$\mu$Jy]} &
  \multicolumn{1}{c|}{AGN} \\
\hline
  1 &                    &                   &                    &                    & 649.83 $\pm$ 25.48 &          &          & \\
  2 & 982.65 $\pm$ 4.09  & 642.35 $\pm$ 3.17 & 463.57 $\pm$ 9.7   & 786.24  $\pm$ 8.26 & 522.05  $\pm$ 12.92 &         &          & \\
  3 & 1601.7 $\pm$ 4.57  & 1112.9 $\pm$ 4.39 & 760.76 $\pm$ 10.71 & 602.03 $\pm$ 5.61 &                     &          &          & \\
  4 & 4056.2 $\pm$ 13.12 & 2806.7 $\pm$ 6.52 & 2756.1 $\pm$ 21.18 & 8360.2 $\pm$ 14.44 & 7091.1 $\pm$ 21.25 & 115570.0 & 574590.0 & \\
  5 &                    &                   &                    &                    & 4851.0 $\pm$ 21.13 & 15020.0  &          & \\
  6 & 1707.0 $\pm$ 6.56  & 1213.2 $\pm$ 3.69 & 1483.1 $\pm$ 17.45 & 5734.6 $\pm$ 11.03 & 9659.0 $\pm$ 21.62 & 129220.0 & 218440.0 & \\
  7 &                    & 1119.1 $\pm$ 7.12 &                   & 514.31 $\pm$ 14.09 & 407.65 $\pm$ 16.67 &           &          & \\
  8 &                    &                   &                   &                    & 1964.0 $\pm$ 14.73 &           &          & \\
  9 &                    & 826.02 $\pm$ 4.27 &                   & 760.54 $\pm$ 12.53 & 1099.9 $\pm$ 16.91 &           &          & \\
  10 &                   &                   &                   &                    & 5237.6 $\pm$ 20.65 & 92560.0   & 360010.0 & \\
  11 &                   &                   &                   &                    & 480.31 $\pm$ 15.1 &            &          & \\
  12 & 1321.2 $\pm$ 2.47 & 924.4  $\pm$ 2.53 & 925.13 $\pm$ 7.79 & 4547.6 $\pm$ 8.6  & 5377.6   $\pm$ 20.53 & 67840.0  & 80250.0  & \\
  13 & 1202.8 $\pm$ 2.89 & 825.9  $\pm$ 3.01 & 515.12 $\pm$ 9.01 & 699.14 $\pm$ 8.67 & 169.76  $\pm$ 17.27 &           &          & \\
  14 & 1134.7 $\pm$ 2.58 & 751.64 $\pm$ 2.74 & 559.29 $\pm$ 6.58 & 1201.9 $\pm$ 7.39 & 932.88 $\pm$ 17.15 &            &          & \\
  15 & 1379.4 $\pm$ 3.89 & 1008.8 $\pm$ 4.39 & 698.5 $\pm$ 8.38 & 1753.8 $\pm$ 8.52 & 2398.5  $\pm$ 16.42 & 1620.0     &          & \\
  16 & 2722.7 $\pm$ 3.44 & 1720.6 $\pm$ 3.18 & 1365.3 $\pm$ 8.46 & 1661.1 $\pm$ 7.45 & 1232.4 $\pm$ 13.53 &            &          & \\
  17 & 512.8  $\pm$ 2.4  & 318.82 $\pm$ 2.88 & 231.09 $\pm$ 7.23 & 182.6 $\pm$ 8.5 & 171.23   $\pm$ 16.42 &            &          & \\
  18 & 929.66 $\pm$ 2.98 & 681.77 $\pm$ 3.56 & 502.83 $\pm$ 8.02 & 1214.6 $\pm$ 9.39 & 1153.0 $\pm$ 17.27 & 6960.0     & 2010.0   & X \\
  19 &                   &                   &                   &                    & 2803.6 $\pm$ 19.2 & 32840.0    &          & \\
  20 & 1653.2 $\pm$ 2.66 & 1281.6 $\pm$ 3.46 & 1343.0 $\pm$ 8.09 & 8447.7 $\pm$ 12.28 & 9574.3 $\pm$ 22.22 & 131310.0  & 396400.0 & \\
\hline\end{tabular}
\end{table*}

\subsection{X-ray, radio and Mid-IR selected type 2 AGN}

The next sample, hereafter COSMOS sample, consists of AGN selected among X-ray and
radio emitters in the COSMOS field \citep{trump07}. Among the 466 objects for which Magellan spectroscopy
was carried out, were 135 type 2 AGN and hybrids (type 2s with a galaxy component; for details 
on the classification see \citealt{trump07}) with optical and IRAC
counterparts. For the optical/near-IR photometry we used the COSMOS first release optical
and near-IR catalogue \cite{capak07}. The IRAC and MIPS photometry was taken from the
S-COSMOS May 2007 release, described in \citealt{sanders07}. 
From the 135 objects, 92 have a 24 micron counterpart and they alone will be the objects
to be studied, as will be discussed in Section \ref{sec:type2s}.

To construct the
SED of these 92 objects we used the $u^{*}$- and $i^{*}$-band CFHT data, the $g^{+}$,
$V_j$, $r^{+}$, and $z^{+}$ Subaru data, the $K_S$ KPNO or CTIO data and the IRAC and MIPS
S-COSMOS data. Photometry from the MIPS 24\mums deep scans was used whenever available.
The $B_j$ band data were discarded due to the large overlap with $g^{+}$-band that would lead
to overly redundant information and the
CFHT $i^{*}$ was used instead of the Subaru $i^{+}$ because some of the objects were missing 
from the Subaru $i^{+}$-band data (even though, according to \citealt{capak07} the latter
observations are deeper by two magnitudes).
Tables \ref{tab:cosmosObj} and \ref{tab:spitzerFluxes} show the COSMOS names, coordinates,
redshifts and type (``q2'' for type 2 AGN, ``q2e'' for hybrid and a question mark next to the
type denotes a questionable classification) taken from \cite{trump07}, Table 2, and the Spitzer 
fluxes (from \citealt{sanders07}), respectively, for the first 20 objects
of the sample. The full tables are available online.

\begin{table*}
\caption{COSMOS names, coordinates, redshifts and classification type for the
COSMOS type 2 AGN sample, from Trump et al. 2007. The full catalogue is available online.}
\label{tab:cosmosObj}
\begin{tabular}{|r|l|l|l|r|r|}
\hline
  \multicolumn{1}{|c|}{SqNr} &
  \multicolumn{1}{c|}{COSMOS Names} &
  \multicolumn{1}{c|}{RA} &
  \multicolumn{1}{c|}{Dec} &
  \multicolumn{1}{c|}{$z$} &
  \multicolumn{1}{c|}{Type} \\
\hline
  1 & COSMOS\_J100033.08+015414.4 & 10:00:33.08 & +01:54:14.4 & 1.16908 & q2?\\
  2 & COSMOS\_J095945.18+023439.4 & 09:59:45.18 & +02:34:39.4 & 0.47717 & q2e\\
  3 & COSMOS\_J095943.76+022008.0 & 09:59:43.76 & +02:20:08.0 & 0.92998 & q2e\\
  4 & COSMOS\_J100035.25+015726.0 & 10:00:35.25 & +01:57:26.0 & 0.84845 & q2e\\
  5 & COSMOS\_J100059.47+022631.8 & 10:00:59.47 & +02:26:31.8 & 0.66944 & q2e\\
  6 & COSMOS\_J100204.84+013356.0 & 10:02:04.84 & +01:33:56.0 & 0.09815 & q2\\
  7 & COSMOS\_J100013.73+021221.3 & 10:00:13.73 & +02:12:21.3 & 0.18645 & q2\\
  8 & COSMOS\_J095933.79+014906.3 & 09:59:33.79 & +01:49:06.3 & 0.13327 & q2\\
  9 & COSMOS\_J100028.69+021745.3 & 10:00:28.69 & +02:17:45.3 & 1.03911 & q2?\\
  10 & COSMOS\_J100107.22+020115.8 & 10:01:07.22 & +02:01:15.8 & 0.24672 & q2\\
  11 & COSMOS\_J100223.62+021132.8 & 10:02:23.62 & +02:11:32.8 & 0.12262 & q2e\\
  12 & COSMOS\_J100211.38+015633.2 & 10:02:11.37 & +01:56:33.2 & 0.21752 & q2\\
  13 & COSMOS\_J100243.25+020521.9 & 10:02:43.25 & +02:05:21.9 & 0.21398 & q2\\
  14 & COSMOS\_J100022.94+021312.7 & 10:00:22.94 & +02:13:12.7 & 0.18583 & q2\\
  15 & COSMOS\_J100135.22+023109.1 & 10:01:35.23 & +02:31:09.1 & 0.21898 & q2\\
  16 & COSMOS\_J095934.89+015241.1 & 09:59:34.89 & +01:52:41.1 & 0.13333 & q2\\
  17 & COSMOS\_J100152.90+020113.7 & 10:01:52.90 & +02:01:13.7 & 0.22042 & q2\\
  18 & COSMOS\_J100118.97+020822.5 & 10:01:18.97 & +02:08:22.5 & 0.16768 & q2\\
  19 & COSMOS\_J100231.87+015926.7 & 10:02:31.87 & +01:59:26.7 & 0.80314 & q2\\
  20 & COSMOS\_J095945.18+015330.4 & 09:59:45.18 & +01:53:30.4 & 0.21987 & q2e\\
\hline\end{tabular}
\end{table*}

\begin{table*}
\caption{COSMOS Spitzer fluxes of the sample presented in Table \ref{tab:cosmosObj}.
The errors on the 70 and 160 \mums fluxes are 10\% of the fluxes.
The full sample is available online.}
\label{tab:spitzerFluxes}
\begin{tabular}{|r|r|r|r|r|r|r|r|}
\hline
  \multicolumn{1}{|c|}{SqNr} &
  \multicolumn{1}{c|}{S$_{3.6}$ [$\mu$Jy]} &
  \multicolumn{1}{c|}{S$_{4.5}$ [$\mu$Jy]} &
  \multicolumn{1}{c|}{S$_{5.8}$ [$\mu$Jy]} &
  \multicolumn{1}{c|}{S$_{8.0}$ [$\mu$Jy]} &
  \multicolumn{1}{c|}{S$_{24}$  [$\mu$Jy]} &
  \multicolumn{1}{c|}{S$_{70}$  [$\mu$Jy]} &
  \multicolumn{1}{c|}{S$_{160}$ [$\mu$Jy]} \\
\hline
  1 & 148.19 $\pm$ 0.26 & 145.3 $\pm$ 0.27 & 108.12 $\pm$ 0.74 & 680.25 $\pm$ 1.55 & 2819.9 $\pm$ 115.8 & 48550.0 & \\
  2 & 329.74 $\pm$ 0.4 & 239.16 $\pm$ 0.42 & 155.24 $\pm$ 0.74 & 302.6 $\pm$ 1.77 & 502.5 $\pm$ 11.5 &  & \\
  3 & 116.34 $\pm$ 0.23 & 102.27 $\pm$ 0.26 & 96.91 $\pm$ 0.77 & 125.99 $\pm$ 1.47 & 562.8 $\pm$ 10.8 & & \\
  4 & 108.87 $\pm$ 0.22 & 114.63 $\pm$ 0.27 & 75.65 $\pm$ 0.74 & 235.19 $\pm$ 1.52 & 687.8 $\pm$ 111.5 & & \\
  5 & 10.37 $\pm$ 0.1 & 18.13 $\pm$ 0.17 & 34.93 $\pm$ 0.61 & 92.23 $\pm$ 1.3 & 851.0 $\pm$ 108.2 & & \\
  6 & 806.09 $\pm$ 0.89 & 627.66 $\pm$ 0.7 & 767.83 $\pm$ 1.38 & 5846.26 $\pm$ 4.76 & 8750.9 $\pm$ 170.7 & 187800.0 & \\
  7 & 132.39 $\pm$ 0.2 & 106.82 $\pm$ 0.25 & 71.29 $\pm$ 0.62 & 262.16 $\pm$ 1.44 & 3295.0 $\pm$ 42.3 & & \\
  8 & 376.01 $\pm$ 0.45 & 279.86 $\pm$ 0.39 & 267.35 $\pm$ 0.84 & 2175.11 $\pm$ 2.45 & 3684.0 $\pm$ 115.0 & 73760.0 & 219900.0\\
  9 & 33.02 $\pm$ 0.14 & 41.25 $\pm$ 0.2 & 43.11 $\pm$ 0.72 & 44.6 $\pm$ 1.41 & 396.7 $\pm$ 112.0 & & \\
  10 & 73.22 $\pm$ 0.17 & 80.59 $\pm$ 0.24 & 56.14 $\pm$ 0.67 & 300.27 $\pm$ 1.58 & 517.2 $\pm$ 96.7 & & \\
  11 & 365.01 $\pm$ 0.4 & 285.97 $\pm$ 0.38 & 255.93 $\pm$ 0.77 & 1464.55 $\pm$ 1.97 & 1957.6 $\pm$ 108.3 & 55820.0 & \\
  12 & 80.37 $\pm$ 0.17 & 75.51 $\pm$ 0.21 & 48.93 $\pm$ 0.61 & 292.38 $\pm$ 1.39 & 654.9 $\pm$ 112.3 &  & \\
  13 & 242.32 $\pm$ 0.33 & 228.27 $\pm$ 0.31 & 172.33 $\pm$ 0.78 & 1058.68 $\pm$ 1.69 & 2214.2 $\pm$ 94.0 &  & \\
  14 & 246.44 $\pm$ 0.31 & 203.64 $\pm$ 0.33 & 159.9 $\pm$ 0.73 & 1273.79 $\pm$ 2.03 & 6399.4 $\pm$ 111.3 & 63680.0 & \\
  15 & 90.39 $\pm$ 0.19 & 84.12 $\pm$ 0.22 & 57.78 $\pm$ 0.68 & 373.63 $\pm$ 1.48 & 1189.4 $\pm$ 222.1 & & \\
  16 & 196.56 $\pm$ 0.27 & 164.13 $\pm$ 0.3 & 136.25 $\pm$ 0.67 & 489.73 $\pm$ 1.55 & 1886.0 $\pm$ 108.8 & & \\
  17 & 118.25 $\pm$ 0.23 & 114.71 $\pm$ 0.27 & 73.56 $\pm$ 0.71 & 454.65 $\pm$ 1.62 & 1164.5 $\pm$ 111.0 & & \\
  18 & 285.06 $\pm$ 0.4 & 231.32 $\pm$ 0.36 & 178.53 $\pm$ 0.84 & 1378.11 $\pm$ 2.13 & 2221.6 $\pm$ 109.6 & 42480.0 & \\
  19 & 22.08 $\pm$ 0.12 & 19.33 $\pm$ 0.18 & 18.82 $\pm$ 0.59 & 38.77 $\pm$ 1.34 & 1843.1 $\pm$ 94.6 & & \\
  20 & 81.2 $\pm$ 0.17 & 72.98 $\pm$ 0.2 & 42.49 $\pm$ 0.67 & 142.2 $\pm$ 1.21 & 432.0 $\pm$ 88.7 & & \\
\hline\end{tabular}
\end{table*}

We also use a list of type 2 AGN taken from the mid-IR selected spectroscopic sample of
galaxies and AGN from the 15\mums ELAIS-SWIRE survey, 
presented in \cite{gruppioni08}. From the 203 spectroscopically identified 
objects in the sample (hereafter ELAIS sample), we selected a total of 23 type 2 AGN
that had sufficient photometric coverage. The ELAIS names and Spitzer (SWIRE) fluxes
for these objects can be found in \cite{gruppioni08}, Table 1.

\subsection{Combined sample properties}
\label{sec:dataall}

\begin{figure*}
\centerline{
\psfig{file=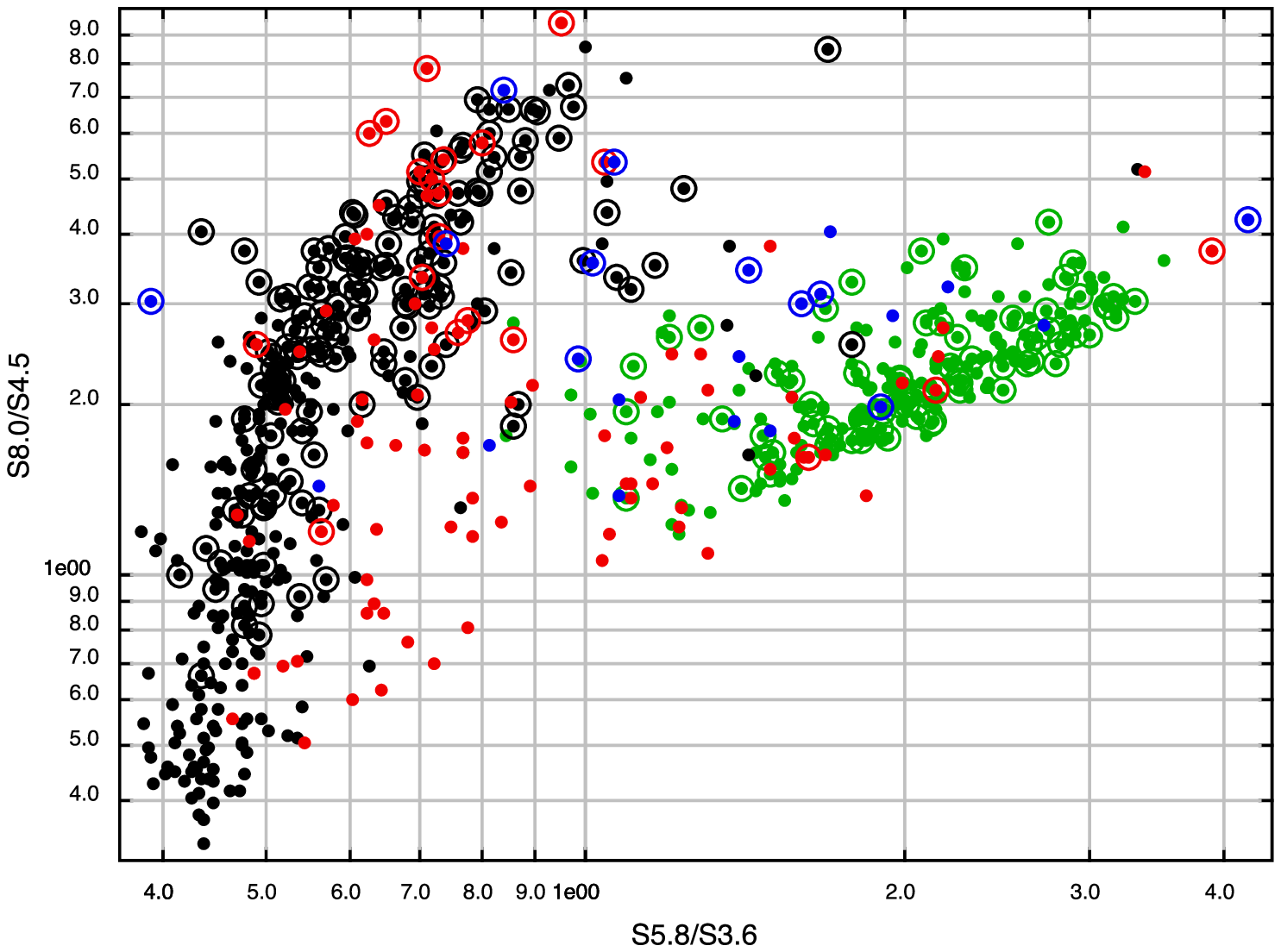, width=10cm}
\psfig{file=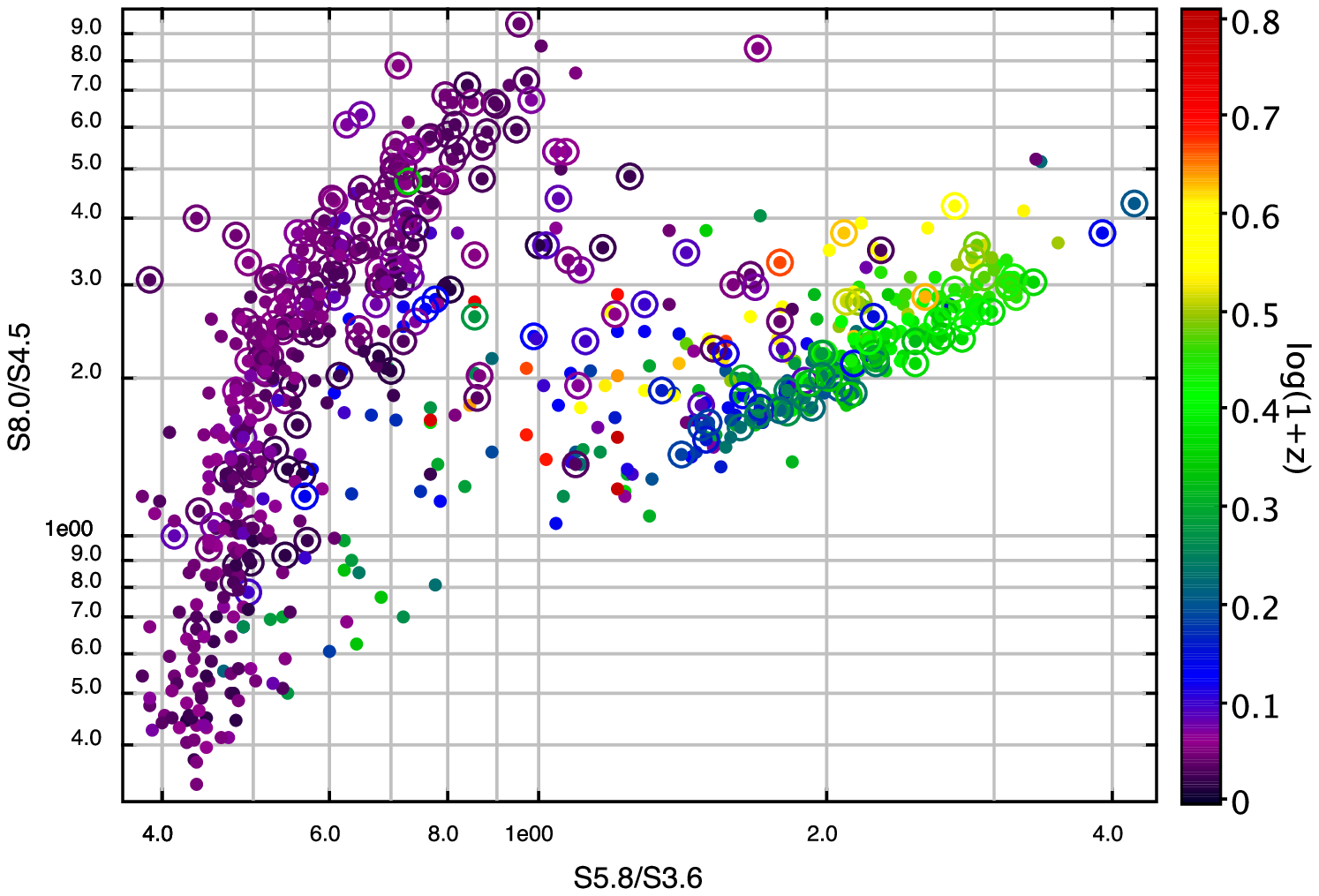, width=10.5cm}}
\caption{IRAC colour-colour plot (S$_{8.0}$/S$_{4.5}$ versus S$_{5.8}$/S$_{3.6}$) 
illustrating the objects of all the samples detected in the four IRAC
bands. Left panel: the samples in black, red, and blue are the low-$z$, COSMOS,
and ELAIS, respectively. The quasar sample (green) is shown for
comparison. The points inside
open circles denote detections at 70\mum. The right panel shows the same 
diagram but with the colours of the points indicating their redshift.}
\label{fig:colours}
\end{figure*}

Fig. \ref{fig:colours} shows the position of the galaxies on the IRAC
colour-colour diagram S$_{8.0}$/S$_{4.5}$ versus 
S$_{5.8}$/S$_{3.6}$. In the left
panel, the samples in black, red, blue, and green are the low-$z$,
COSMOS, ELAIS and the quasar samples, respectively. The points inside
open circles denote detections at 70\mum.
The different loci formed by the four different samples reflect the distinct
selection techniques and hence properties of the objects.
As suggested by \cite{sajina05} from a study based on empirical templates,
and as will be demonstrated shortly, the objects at the lower
left of the plot are dominated by stellar emission;
the objects rising in S$_{8.0}$/S$_{4.5}$ while maintaining
a low S$_{5.8}$/S$_{3.6}$ (i.e., the
bulk of the low-$z$ sample with about a third of the COSMOS sample lying in
the same region) are those dominated by PAH emission 
while the objects with both colours rising,
towards the upper right corner of the plot, are continuum
dominated objects. The quasar locus
occupies a very well confined region of the colour space, as
shown already (e.g. \citealt{lacy04});
in their majority and up to a redshift of $\sim$2, they lie on a straight line of 
slope of one, simply indicating the rise of the torus emission as a power law.
We should note that the position of X-ray selected AGN has already been shown
by \cite{cardamone08}, that sample however contained a much smaller
fraction of objects with star formation, and was plagued by a large fraction of
objects with unknown redshifts. Interestingly, their colors tend to populate
the lower right part of the
colour space, which is strikingly empty in our larger sample.

\begin{figure}
\centerline{
\psfig{file=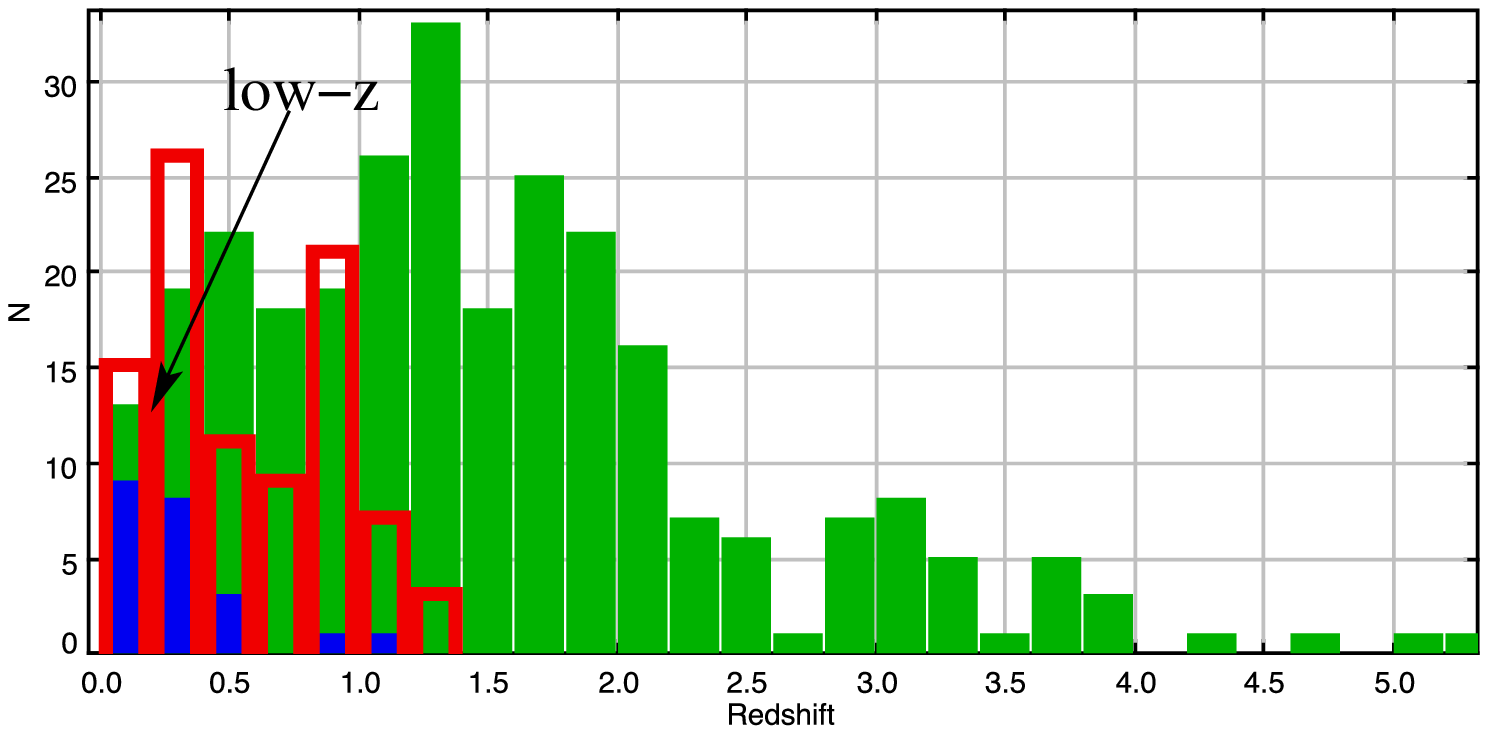, width=10cm}}
\centerline{
\psfig{file=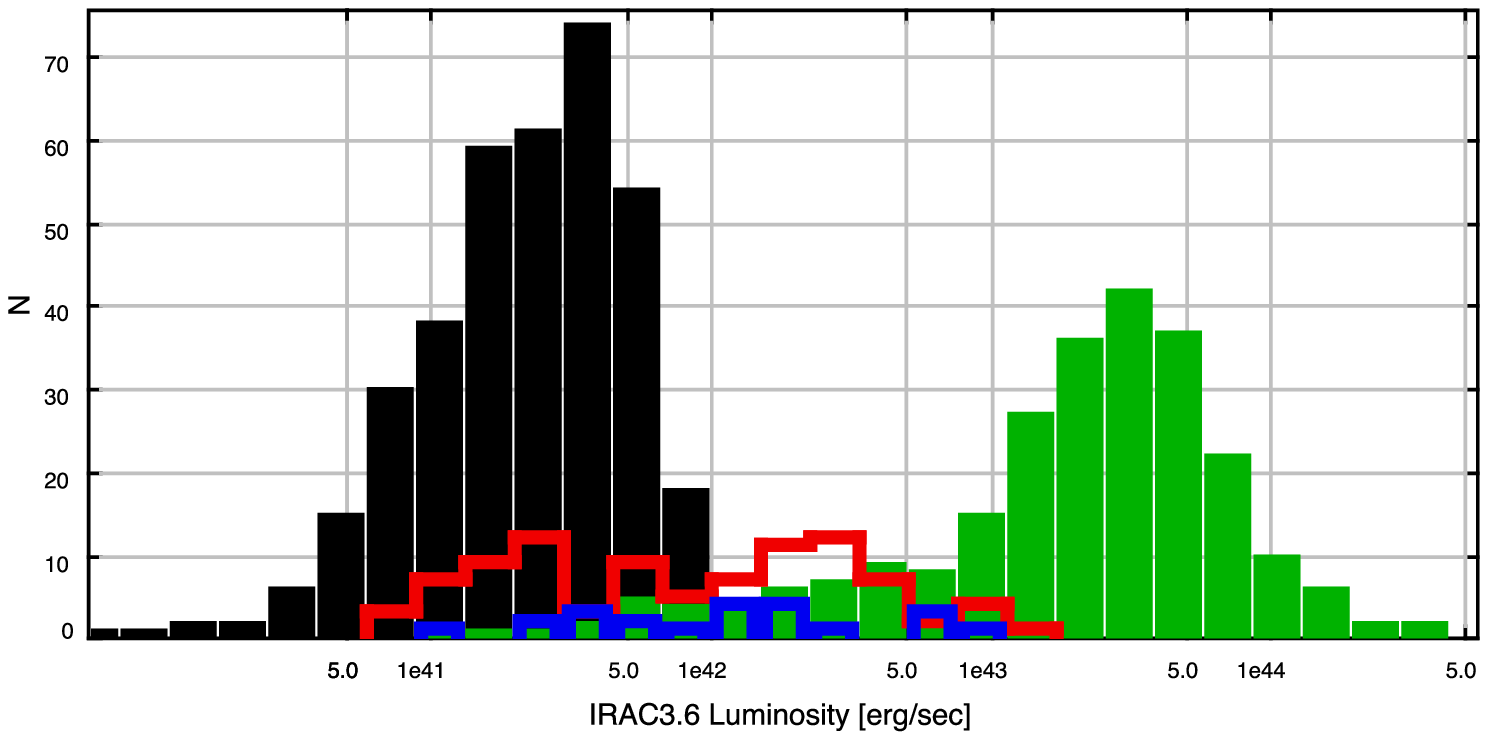, width=10cm}}
\centerline{
\psfig{file=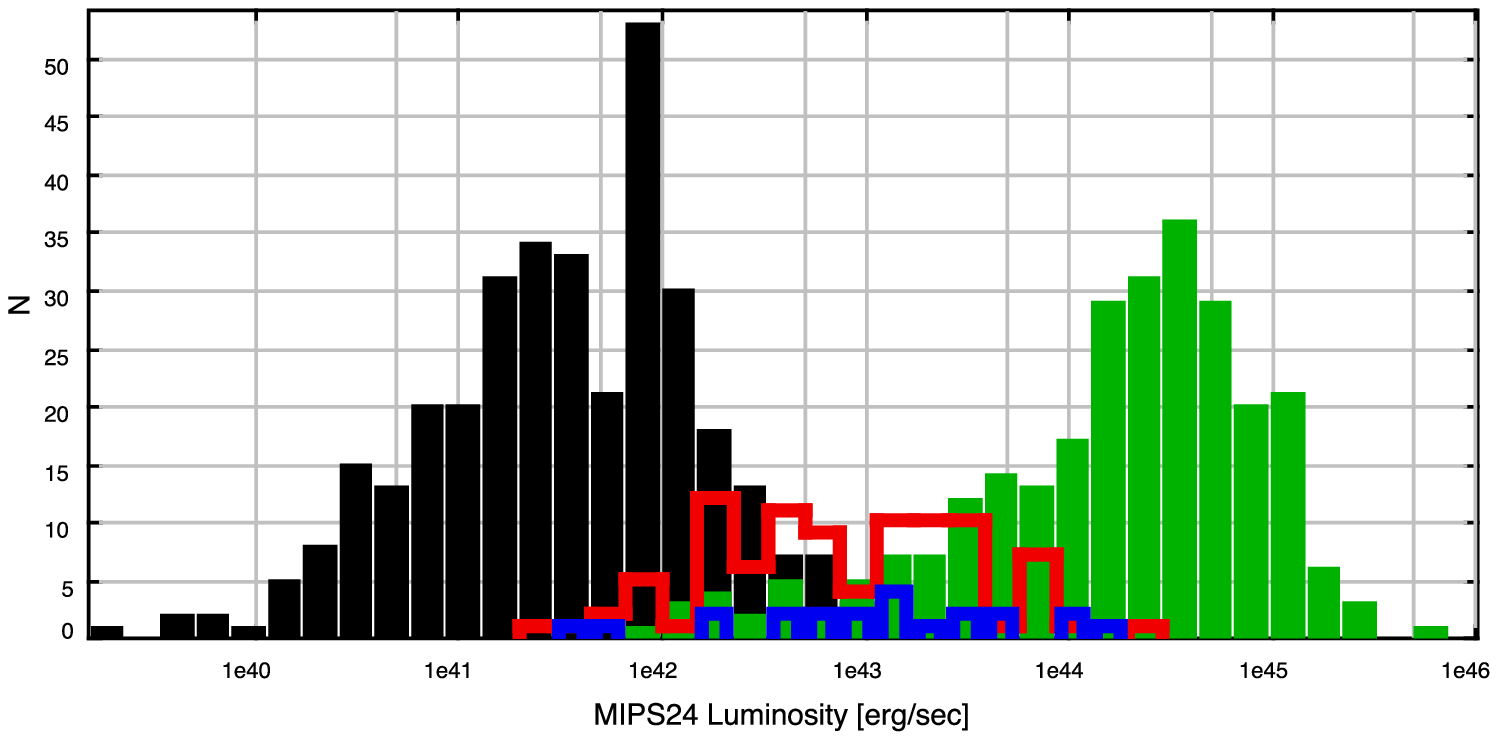, width=10cm}}
\caption{Redshift (upper panel), 3.6 \mums luminosity (middle panel) and 24 \mums luminosity 
(lower panel) histograms for for the low-$z$, COSMOS, ELAIS and quasar samples.
Colour coding follows Fig. \ref{fig:colours}.  The black arrow indicates the location
of the low-$z$ sample.
}
\label{fig:zhisto2}
\end{figure}

One should keep in mind that the
various samples have not only different observed properties
but also very different average redshifts, as seen in the right
panel of Fig. \ref{fig:colours}. A histogram of the redshift
distributions is shown in the upper panel of Fig. \ref{fig:zhisto2}.
The middle and lower panels of Fig. \ref{fig:zhisto2}
show the distributions of the 3.6 and 24 \mums luminosities of the various
samples. 

\section{OBSERVED AND MODEL SEDS}
\label{sec:seds}

The observed UV to FIR SED of a galaxy can be decomposed in three distinct components:
stars with the bulk of their power emitted in the optical and near-IR, hot dust
mainly heated by UV/optical emission from gas accreting 
onto the central supermassive black hole and whose emission
peaks somewhere between a few and a few tens of microns, and cold dust principally heated by
star formation. In the present work we consider all three components, which we model as 
follows.

\subsection{The stellar component}
\label{sec:stars}

The stellar component is the sum of Simple Stellar Population (SSP) 
models of different age, all having a common (solar) metallicity. The set of SSPs
is built using the Padova evolutionary tracks \citep{bertelli94}, a Salpeter
IMF with masses in the range
0.15 -- 120 M$_\odot$ and the \cite{jacoby84} library of observed 
stellar spectra in the optical domain. The extension to the UV and IR range is derived
from the Kurucz theoretical libraries. Dust emission from circumstellar envelopes 
of AGB stars has been added by \cite{bressan98}. The weight in the final spectrum of each SSPs,
as a function of age, is computed according to a Schmidt law for the star formation rate:
\begin{equation}\label{eqn:schmidt}
SFR(t)=\left(\frac{T_G-t}{T_G}\right)\times \exp\left(-\frac{T_G-t}{T_G\cdot\tau_{sf}}\right)
\end{equation}
where $T_G$ is the age of the galaxy (i.e. of the oldest SSP), which is assumed to be as old as
the age of the universe at the galaxy's redshift, and $\tau_{sf}$ is the 
duration of the burst in units of $T_G$. Extinction is applied to the 
final SED by assuming a uniform foreground dust screen with a standard Galactic 
extinction law \citep{cardelli89}. The free parameters are, therefore, the duration of the initial burst and the 
amount of extinction.

\subsection{The torus component}
\label{sec:tori}

The models for dust emission in AGN are those presented in \citealt{fritz06}.
To summarize, the dust distribution is smooth, the torus geometry is that of a ``flared disk'',
the dust consists of graphite and silicate grains. The graphite grains, with 
sublimation temperature around 1500 K are responsible for the black body like
emission in the near-IR part of the AGN spectra, creating a minimum in the
observed SEDs between the falling accretion emission and the rising torus
emission, while the silicate grains are responsible for the absorption feature
at 9.7 micron, seen in the IR spectra of all type 2 AGN.
The size of the inner torus radius, $R_{in}$, depends both on the sublimation 
temperature of the grains, $T_{1500}$, and on the accretion luminosity, 
$L_{acc}$, according to \cite{barvainis87}:
\begin{equation}
\label{eqn:rin}
R_{in}\simeq 1.3 \cdot \sqrt{\rm L_{acc}}\cdot T_{1500}^{-2.8} [pc]
\end{equation}
The dust density within the torus is allowed to vary along 
both the radial and the angular coordinates: 
\begin{equation}
\label{eqn:density}
\rm \rho(r,\theta)=\alpha \cdot r^\beta \cdot e^{-\gamma |\cos(\theta)|}
\end{equation}

We give here, as a reminder, the parameter values for the discrete grid of torus models that will 
be used, as presented in Paper 1: $\beta=0.0, -0.5, -1.0$ and $\gamma=0.0, 6.0$ (see Eq. \ref{eqn:rin}),
optical depth $\tau_{9.7}=1.0, 2.0, 3.0, 6.0, 10.0$; torus opening angle $\Theta$=20$^o$, 40$^o$, 60$^o$;
and outer-to-inner radius ratio $R_{out}/R_{in}$=30 and 100. This last set of parameters only holds when $\beta=0.0$,
i.e. when the dust density is constant with the distance from the centre. For the models
with $\beta < 0.0$, $R_{out}$ is re-calculated to be the distance at which the dust density
drops at 10\% of its value at $R_{in}$.

\subsection{The starburst component}
\label{sec:sb}

For the cold dust component, which is the major contributor to the bolometric emission at wavelengths
longer than $\sim 30$ \mum, we choose six well-studied observational SB templates:
M82 as a representative of a ``typical'' SB IR emission, Arp 220
as representative of a very extinguished starburst and the templates of NGC1482,
NGC4102, NGC5253 and NGC7714, as intermediate SB templates.

\subsection{Fitting the observed SEDs}
\label{sec:sedfit}

The SED fitting procedure is that of a standard $\chi^2$ minimisation.
As a first step, the UV to near-IR (rest-frame) data points are fitted by means of 
a stellar host component that dominates the emission in this range
(since the accretion emission of the low luminosity AGN of the sample is most probably
masked under the stellar light). The best fit to the observed data points in this wavelength 
range is found by exploring a grid of $\tau_{sf}$ and $E(B-V)$ values (see Sec. \ref{sec:stars}).

In the next step, the mid-IR points are then fit by a torus component. 
Since all the objects under study are type 2 AGN and therefore the central source
is hidden behind the torus,
we will focus on the results of the run realised with torus models of high optical
depth, $\tau_{9.7} \ge 1.0$ with the exception of the low-$z$ sample,
for which we will allow a full run, including $\tau_{9.7} < 1.0$ torus models, 
since for these low luminosity objects obscuration might also be the result of the
dust in the host galaxy.
In order to properly constrain the 
part of the SED where the cold dust is dominant, one would need at least one FIR measurement,
in this case a 70  and/or 160 \mums detection (the mid-Infrared 24 \mums point alone
is probably not enough to distinguish between a torus-like and a starburst-like emission). 
Correspondingly, an SB component will only be allowed in the presence of a 70
(and/or 160) \mums datapoint(s), along the lines of Paper 1. A single deviation from this
will be discussed in Section \ref{sec:samplek}.

As detailed in Paper 1, the SED fitting leads to the computation of a series of physical 
parameters, namely the accretion luminosity, $L_{acc}$, the IR luminosity, $\rm L_{IR}$,
integrated between 1 and 1000 \mum,
the inner and outer torus radii, $R_{in}$ (defined in Eq. \ref{eqn:rin}) and $R_{out}$, 
the optical depth (or extinction)
and hydrogen column density along the line of sight, the covering factor, CF, i.e. the fraction
of the radiating source covered by the obscuring material, and the mass of
dust, $M_{Dust}$. A quantity that was not dealt with in Paper 1 due to the nature of the sample
(type 1 quasars and hence almost complete lack of stellar components) is the stellar mass,
$M^*$, derived directly by the SSPs, that we will also include in the analysis of the low-$z$ sample.

For the degeneracies related to the various torus models parameters and
the limitations inhibited in our approach, we defer the reader to sections 5.6 and 6 of Paper 1,
where all these issues were discussed in detail. Dealing with quasars, the stellar and SB components were of 
secondary importance. In the present study, however, these two components play
a more important role. The stellar component (SSPs), even though used to fit mainly the UV-to-optical
part of the SEDs whenever necessary, is not studied any further and therefore discussing
degeneracies between the relevant parameters is well outside the scope of this work, but a
relevant discussion can be found in \cite{berta04}. The only remaining issue is that of the
normalisation of the SB template, particularly in the absence of a sufficient number of data points.
After an initial normalisation of the mentioned temlpate to the 70 \mums point, an iterative process
results in the best normalisation factor after the torus emission has been excluded.
In other words, the SB template is finally used to fit the emission that is not accounted for
by the torus model. This could introduce a bias, maximising the estimated AGN and minimising the
SB contributions, respectively. However, since the iterative process takes into account not only the
best fit torus model but the best 30 torus models, we are confident that these deviations will not be more
than a few percent.

\section{TESTING THE INDIVIDUAL SAMPLES}
\label{sec:fitresults}

\subsection{Low-luminosity, low-redshift AGN}
\label{sec:samplek}

\begin{figure}
\centerline{
\psfig{file=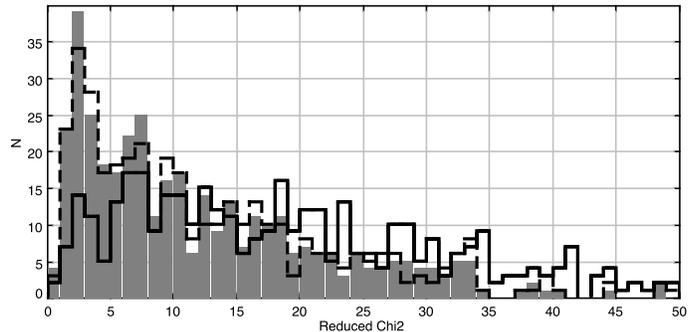,width=10cm}}
\caption{Distribution of the minimum $\chi^2$ from the analysis of the low-$z$ sample
from the ``standard'' run when no SB component was allowed in the absence of 70
(and/or 160) \mums data (plain line) and from the run allowing an additional
SB component in the presence of a 24 \mums detection (dashed line). In gray,
is the final $\chi^2$ distribution selecting, for each object, 
the minimum $\chi^2$ between the two runs.}
\label{fig:KhistoChi2}
\end{figure}

The distribution of the reduced $\chi^2$ of the sample is very broad and in general the values
are quite large (Fig. \ref{fig:KhistoChi2}, plain line). The reasons for these high values has been 
explained in detail in Paper I
and relate, among other things, to the small values of the photometric errors as well as the crudeness
of the X-ray-to-optical spectrum of the AGN models. 

\begin{figure*}
\centerline{
\psfig{file=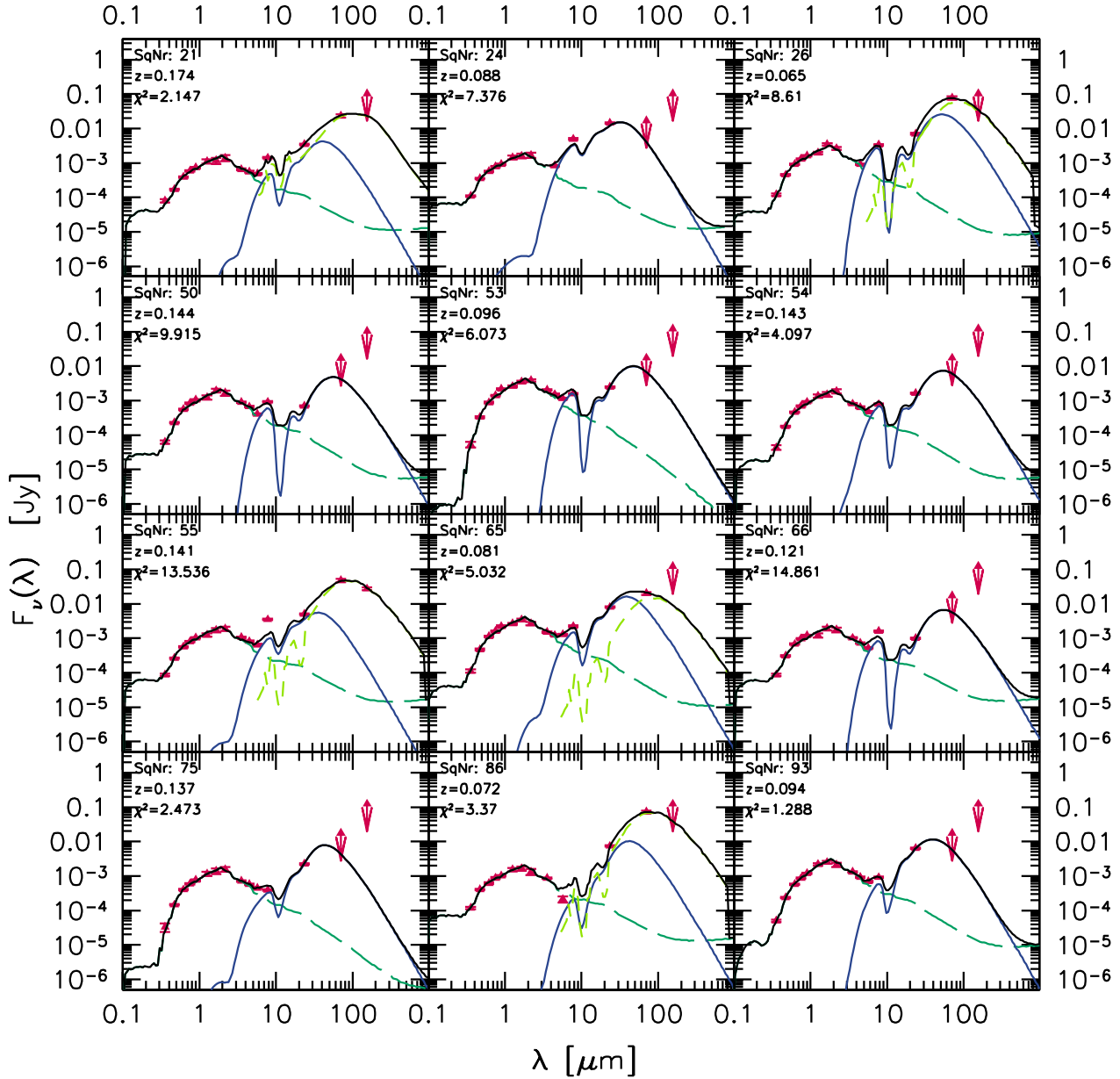,width=.97\textwidth}}
\caption{Best fit SEDs for the first 12 low-$z$ objects with reduced $\chi^2<16.0$ that were assigned a
torus component. Data points (in the rest frame) are shown in red; the stellar model denoted with a green dashed line, 
the torus with a blue line, and the cold dust component with a dashed light green line.}
\label{fig:kaufffits}
\end{figure*}

Examples of the observed SEDs and the best fit models for objects assigned 
a torus template are shown in Fig. \ref{fig:kaufffits}. The sequence number 
(SqNr) shown in the plots correspond to the SqNr shown in table \ref{tab:sdssObj}. 
The signature of a torus would be a large increase of the 24 \mums
flux with respect to the 8.0 \mums point while the presences of a starburst 
will usually be revealed by a rise in the 8 \mums flux
due to the PAH features, but with a subsequent flattening or slow rise of the SED 
towards the redder (24 \mum) wavelength. The absence of any component other 
than the stellar is marked by a monotone decrease
of the SED after the stellar near-Infrared bump ($\sim 2-3$ \mums).
The best fits for all 388 objects are provided as online material.

\begin{figure*}
\centerline{
\psfig{file=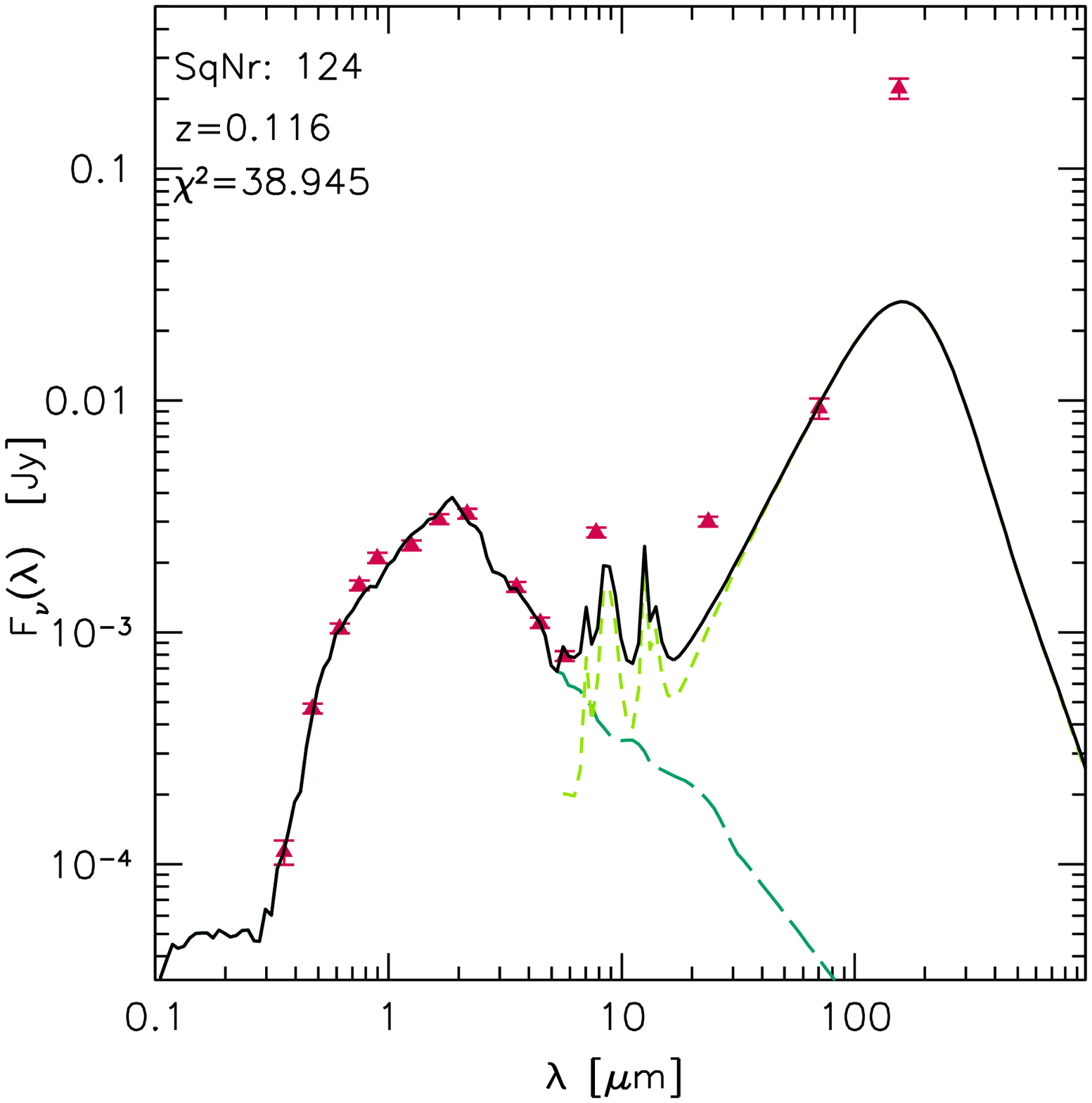,width=.35\textwidth}
\psfig{file=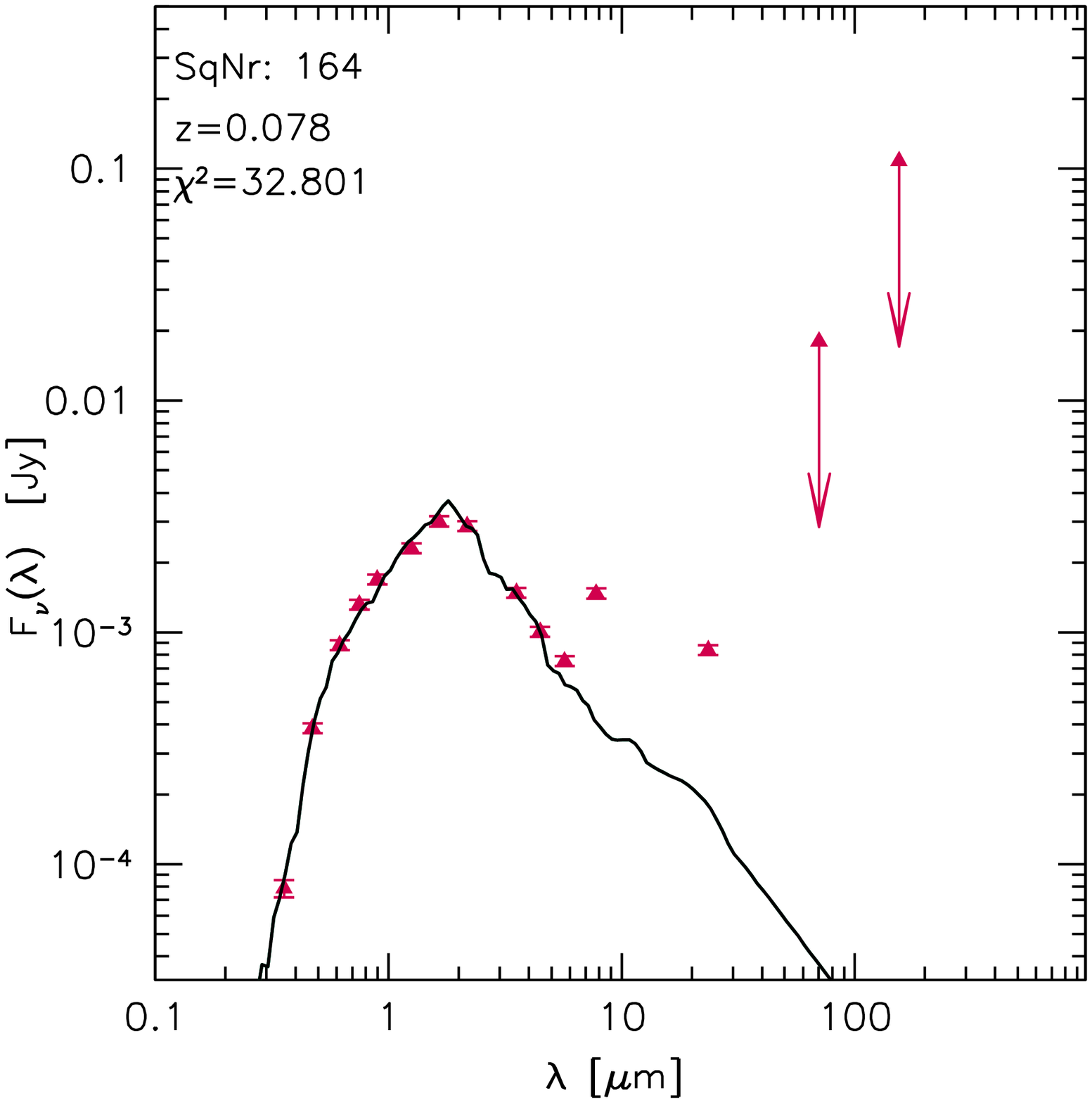,width=.35\textwidth}
\psfig{file=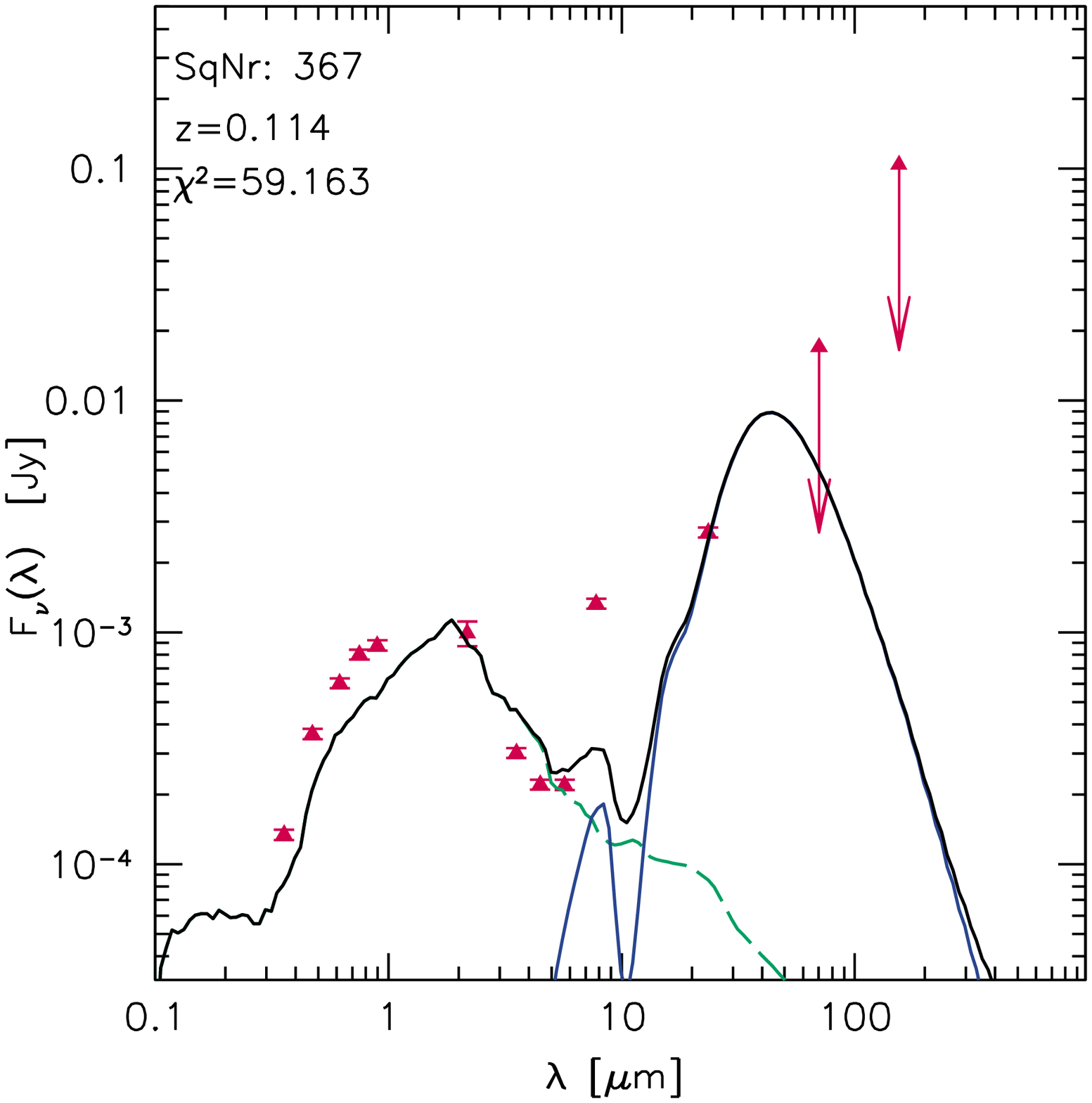,width=.35\textwidth}}
\caption{Examples of failed fits for three objects in the low-$z$ sample.}
\label{fig:badkaufffits}
\end{figure*}

In Fig. \ref{fig:badkaufffits} we also provide three examples of objects for which
the fitting method completely failed to reproduce the observed SEDs, 
in order to give the reader a flavour of the various problems we have encountered.
Object with SqNr 124 (left plot in Fig. \ref{fig:badkaufffits}) has very strong
mid-to-far infrared emission that none of our SB templates was able to account for;
object 164 (middle plot) has a stellar-like SED up to 5.8 \mum, but the subsequent
rise in flux at 8 \mums and drop at 24 \mums can not be reproduced by any combination
of torus and/or SB templates; and the fitting method simply did not work for object Nr
367 (right plot), where both the assigned SSPs and the torus model reproduce very poorly the observed
SED.

\begin{figure}
\centerline{
\psfig{file=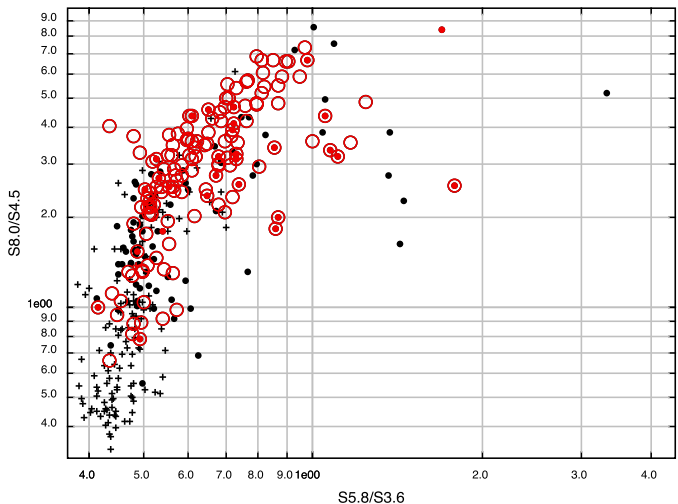,width=10cm}}
\caption{IRAC colour-colour diagram (see Fig. \ref{fig:colours}) indicating the objects
with a stellar component alone (stars), with a torus component (small filled circles)
and a starburst component (large open circles). The symbols in red represent objects with a
70 \mums cold dust component.}
\label{fig:irackauff}
\end{figure}

Going back to the best fits now,
there is a correspondence between the weighted contribution of the various model components
and the IRAC colours of the objects. Fig. \ref{fig:irackauff} illustrates
the position of the objects on the IRAC colour diagram (as in Fig. \ref{fig:colours}).
In agreement with the predictions (e.g. \citealt{sajina05}) almost all objects assigned
a stellar component alone are gathered at the lower left part of the diagram (stars),
starbursts (large open circles) form the bulk of the sample and AGN (objects with a
torus component shown as small filled circles) occupy the space towards the upper right part
of the diagram.  Note however, the large overlap between SBs and stellar-dominated galaxies, which
suggests their close physical relationship and fitting redundancy between
these two types of galaxies.

Drawing a conclusion on the nature of the objects with no 70 \mums detection is actually 
not straight forward. There are cases where the IRAC
fluxes are monotonically decreasing, tracing unmistakably the stellar emission.
While in other objects 
the 8 and/or 24 \mums flux is slowly increasing (unlike a torus signature), 
indicating the presence of a starburst, yet no FIR is detected for some of these cases.
Objects with predominantly, starburst as opposed to AGN activity 
and SEDs dominated by emission in the MIR ($\sim
24$ \mum) rather than the FIR ($\lambda \ge 60$ \mum) are actually known to exist
(see e.g. \citealt{engelbracht08}).

The observed SEDs of these objects were poorly reproduced with the
combination of a stellar and a torus components alone. As a test, and 
conversely to the process followed in Paper I, we performed an extra run allowing
for the use of an SB template even in the absence of a 70 \mums detection.
The results of this run should, therefore, influence the objects without 70 \mums
counterparts, representing however the majority (2/3) of the sample.
The distribution of the reduced $\chi^2$ for this run, shown by a dashed line in 
Fig. \ref{fig:KhistoChi2}, is narrower than that of the ``standard'' run, peaking
at $\chi^2 \sim$ 2. This improvement comes mainly from better fitting the 8 and/or
24 \mums points of the objects that were only assigned a stellar component,
by an additional SB component.
In order to combine the results of the two runs, we will simply select the best $\chi^2$
from the two runs, shown in Fig. \ref{fig:KhistoChi2} with a gray histogram.
For consistency with the study in Paper 1 we will consider ``good'' all fits with
$\chi^2 < 16$, whose limit includes $\sim$70\% of the sample. Note, however, that
we run the same analysis on the entire sample (388 objects) and found
{\bf none} of the results differ in any way if all fits are included.

The sample of objects harbouring an AGN according to the present analysis,
shows very specific trends: the majority of objects
(87\%) were assigned very high optical depth tori ($\tau_{9.7}=10.0$), and 
90\% were assigned tori models with decreasing dust density from the centre ($\beta < 0.0$
in Eq. \ref{eqn:density}). In fact, there is a clear preference for models with $\beta=-0.5$,
with 72\% of the objects assigned such a torus component.
Almost 85\% of the sample have small outer to inner radius ratio ($R_{out}/R_{in}=30$).
The combined result is very small tori with inner radii of less than 1.4 pc (
0.185 pc average) and outer radii reaching in a few cases 100 pc.
These average values
are $\sim$2 
times smaller than the average values for $R_{in}$ 
derived for the quasars and reflect the lower average
accretion luminosity, $L_{acc}$, of the low-$z$ sample. In fact, the lower (by a factor of $\sim$25) 
$\langle L_{acc} \rangle$ of this sample with respect to the quasars is not only due to the
different redshift ranges covered by the two samples but also to the very nature of the
objects.

\begin{figure}
\centerline{
\psfig{file=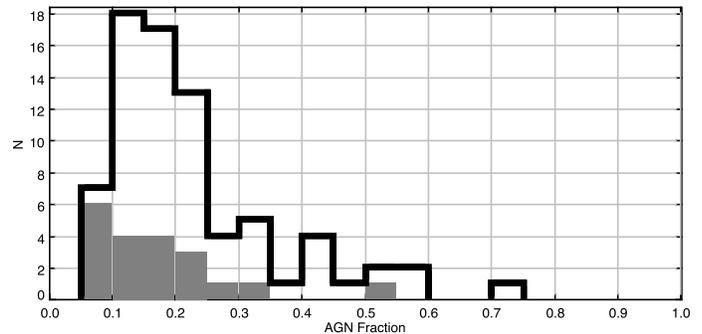, width=10cm}}
\caption{Histogram of the fraction of IR luminosity attributed to the AGN, for objects assigned a torus
component (open histogram) and those with an extra SB component (gray region).}
\label{fig:kauff_agnFrac}
\end{figure}

The contribution of the AGN to the IR luminosity, for the 75 ``good'' fits that have been assigned
a torus component, is shown in Fig. \ref{fig:kauff_agnFrac},
with the grayed-out region corresponding to the 20 objects with an extra SB component.
Due to the nature of the objects 
the AGN fraction tends to be very low compared to the values derived from the 
SWIRE/SDSS quasar sample of Paper 1.

\begin{figure}
\centerline{
\psfig{file=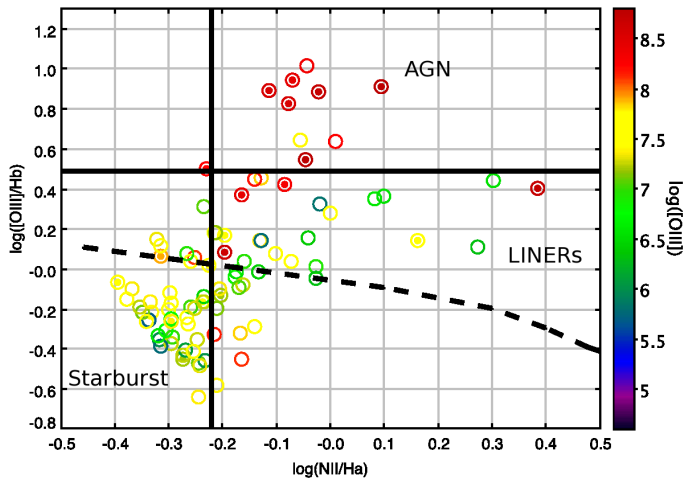,width=10cm}}
\caption{Emission line ratios as a function of extinction corrected {\sc [Oiii]} 
line luminosity, for those objects with a 70 \mums counterpart. Large 
open and small filled circles denote SB and torus component,
respectively. The dashed line shows the usual - and more conservative - AGN
selection criterion suggested by Kewley et al. (2001), according to which the
AGN should occupy the upper right part of the graph.} 
\label{fig:lineRatios}
\end{figure}

All 91 objects with a 70 \mums detection, for which the fits are optimally constrained, 
were assigned a starburst component; of that total, 20 of them show evidence for the presence of a torus.
Fig. \ref{fig:lineRatios} depicts the relative intensities of emission lines (log({\sc [Oiii]}/H$\beta$) 
versus log({\sc [Nii]}/H$\alpha$), as a function of the extinction corrected {\sc [Oiii]} luminosity.
Six of nine objects that occupy the space commonly agreed to be
populated by Seyfert 2 galaxies (upper right corner in Fig. \ref{fig:lineRatios})
have indeed been assigned an AGN component. Another four lie very close to the dividing lines,
and a total of six in the LINERs' region (lower right part). The majority of the objects were assigned an
SB component alone and would have not been part of the AGN catalogue at all,
had the \cite{kewley01} AGN selection criterion been applied (shown here in a dashed line).

The average values for the {\sc [Oiii]} luminosity and the log({\sc [Oiii]}/H$\beta)$ ratio for the objects
with an AGN component, an additional SB component and an SB component alone are given in table
\ref{tab:lineRatios}. Both quantities have larger values when an AGN component is present and lower
when an SB component alone is responsible for the cold(er) dust emission.

\begin{table}
\caption{The average values for the {\sc [Oiii]} luminosity and the log({\sc [Oiii]}/H$\beta)$ ratio for the objects
with an AGN component, an additional SB  component and an SB component alone.}
\begin{tabular}{l r r r}
\hline
\hline
Component & \# & log({\sc [Oiii]}) & log({\sc [Oiii]}/H$\beta$) \\
\hline
\hline
AGN      &  20 & 8.18 $\pm$ 0.60 &  0.299 $\pm$ 0.42 \\
AGN $+$ SB       &  91 & 7.38 $\pm$ 0.75 &  0.006 $\pm$ 0.37 \\
SB alone &  71 & 7.16 $\pm$ 0.64 & -0.077 $\pm$ 0.31 \\
\hline
\end{tabular}
\label{tab:lineRatios}
\end{table}

Our analysis is only able to identify a torus component in about one-third of the total 
low-$z$ sample (75 [142 if no cut in the $\chi^2$ is imposed] out of 266 [388]).
Even though evidence of changing behaviour of (clumpy) tori \citep{hoenig07} or even complete disappearance
of the torus \citep{elitzur06b} towards the lower luminosity regimes ($L < 10^{42}$ erg/sec)
has been reported, these effects are not what we observe here, as the lowest accretion
luminosity estimated for objects with a torus component is of the order of 10$^{44}$ erg/sec,
seen in the first plot of Fig. \ref{fig:compare_props}.
We therefore conclude that we either are unable to detect the signature of AGN with 
$L_{acc}<10^{44}$erg/sec or that the criteria used by \cite{kauffmann03} to select AGN were too generous
and many objects with no nuclear activity may have made it into the suggested AGN sample.

\begin{figure}
\centerline{
\psfig{file=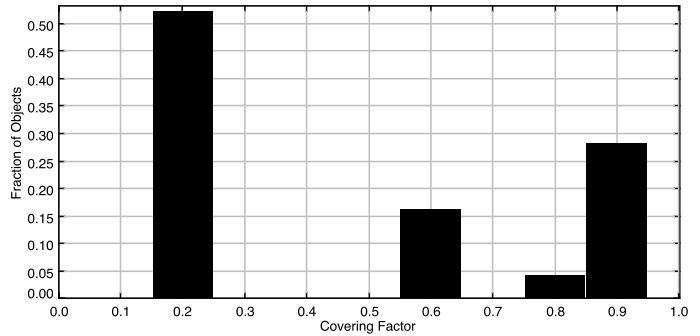,width=10cm}}
\caption{Covering Factor distribution for the low-$z$ sample.}
\label{fig:kauffCF}
\end{figure}

Another issue to consider is the distribution of the resulting covering factor, shown in
Fig. \ref{fig:kauffCF}.
Were the computed values for CF correct, they would translate to a ratio of
type2:type1 objects of about 1:1 in the local and low-$z$ universe, a much
lower value than that observed.

This discrepancy could again be seen as a result of the AGN
candidate selection and the somewhat sparse samples of the observed SEDs,
the combination of which does not always allow the fitting code to give robust results.

Allowing for torus models with $\tau_{9.7}<1.0$ does not have any effect on the results,
as only 18 out of the 388 objects where better fit by a model including such a torus component, 
with a single one among them assigned a $\chi^2<16.0$.

\subsubsection{Properties of the host galaxies}

\begin{figure}
\centerline{
\psfig{file=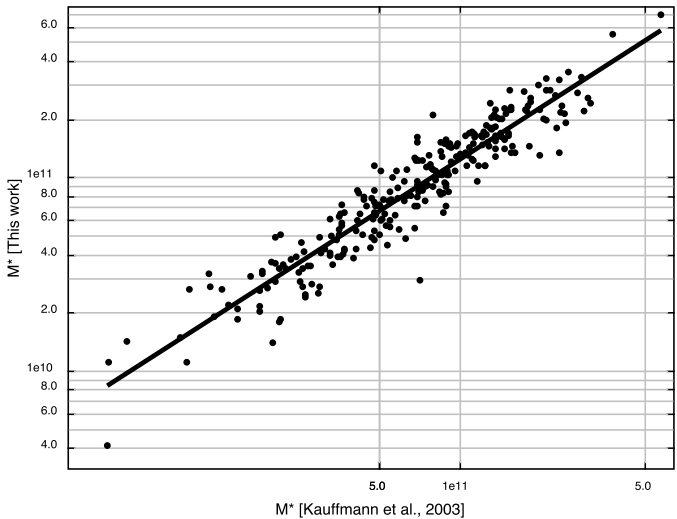,width=10cm}}
\caption{Comparison of the stellar mass derived from this work based on SED fitting 
and that presented in Kauffmann et al. 2003.}
\label{fig:kauff_mstars}
\end{figure}

The SSPs used to reproduce the blue part of the SEDs provide information about the
stellar mass of the host galaxy. To compute the stellar mass we exploit the fact that SSP spectra are
given in luminosity units per one solar mass. We also account for the fact that a certain percentage
of stars, depending on the SSP's age, has evolved and is not shining anymore, so the mass value we
provide is actually ``luminous mass'',
Fig. \ref{fig:kauff_mstars} shows a comparison of the stellar mass computed in this
work and that provided in the narrow line AGN catalogue \cite{kauffmann03}. The line shows
the linear correlation between the two quantities, $M^*_{here}$ = 1.59 + (1.06$M^*_{\bf K}$). Note
that the systematic differences in the mass values are the result of the different IMF used by \cite{kauffmann03},
a Kroupa (2001) IMF, which differs from ours both in the slope and in the mass limits. 

\begin{figure}
\centerline{
\psfig{file=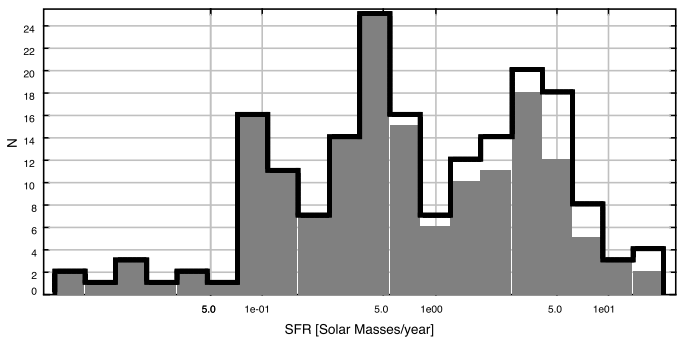,width=10cm}}
\caption{Star formation rates for all objects with an SB component
and $\chi^2 < 16.0$ (black histogram) and those without an AGN component
(gray histogram).}
\label{fig:SFRkauff}
\end{figure}

We now compute the star formation rate (SFR) according to \cite{kennicutt98} as 

\begin{equation}
SFR[M_{\odot}/yr] = 4.5 \times 10^{-44} \times L_{FIR}
\label{eqn:sfr}
\end{equation}

where $L_{FIR}$ is the infrared luminosity of the SB component,
integrated between 8 and 1000 micron. Fig. \ref{fig:SFRkauff}
shows the distribution of the SFRs for all objects with an SB component
and $\chi^2 < 16.0$ (black histogram), with the grayed region corresponding to 
the objects without an AGN component. The average SFR of the entire sample
without imposing a cut on the $\chi^2$ values is of 2.8 $M_{\odot}$/yr, and drops to
2.2 when only objects with $\chi^2 <16.0$ are consider, and to 1.85 when objects
with an AGN component are excluded.

Note that all objects with a torus component are gathered in the part of the histogram with
the higher SRFs. We will get back to the discussion on SFRs in Section \ref{sec:sfr}.

\subsection{Type 2 AGN and hybrids from the COSMOS sample
and Mid-IR selected ELAIS AGN}
\label{sec:type2s}

The spectroscopic sample of X-ray emitters selected in the COSMOS field
presented in \cite{trump07} comprises 80 type 2 AGN, 47 type 2 AGN and red galaxy hybrid
and 8 type 2 AGN whose classification is not 100\% secure. 
Ninety two out of these 135 objects have a 24 micron counterpart and, as already mentioned in 
Section \ref{sec:data}, we will be concentrating on
this sub-sample alone. This was not a constraint applied to the low redshift sample, however,
due to the average (low) redshift of this sample as well as its nature (see Fig. \ref{fig:zhisto2}).
The SEDs are dominated by the stellar light all the way to the IRAC3 band (5.8 \mum)
and even though the presence of a torus is somewhat revealed from the lower S$_{8.0}$/S$_{5.8}$
colour (with respect to objects without a torus component), a single data point at 8.0 \mums
(IRAC4) is not enough to constrain the torus properties. For the record,
36 of the remaining objects (those without a 24 \mums detection) were indeed assigned a torus
component while six others were better represented by SSPs alone.

\begin{figure}
\centerline{
\psfig{file=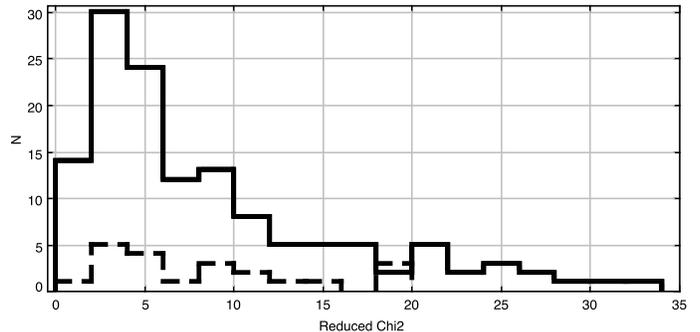,width=10cm}}
\caption{Histogram of the minimum $\chi^2$ from the analysis of
the COSMOS (solid line) and the ELAIS (dashed line) samples.}
\label{fig:TGhistoChi2}
\end{figure}

Fig. \ref{fig:TGhistoChi2} shows the minimum $\chi^2$ distribution for the sub-sample
of 92 objects whose SEDs extend at least up to 24 \mum,
(solid line), excluding two objects with minimum $\chi^2>90$ that are
obviously erroneous and will not be taken into account in what follows.
Again for consistency, we will adopt a cut-off $\chi^2$ of 16, keeping in mind though that
including the rest of the fits will not influence the results in any significant way.

\begin{figure*}
\centerline{
\psfig{file=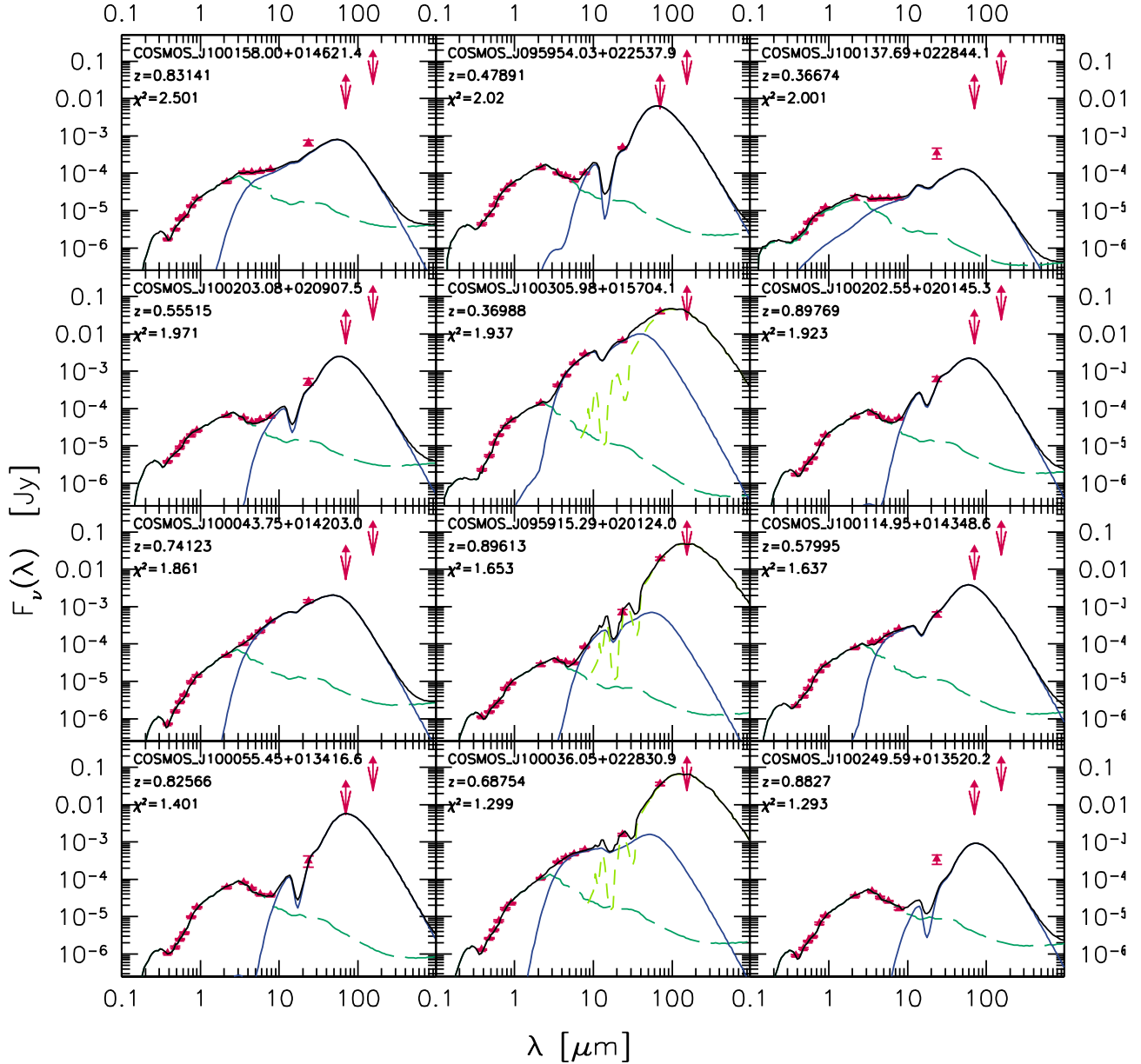,width=.97\textwidth}}
\caption{The best fits for the first 12 COSMOS objects with reduced $\chi^2<16.0$ that were assigned a
torus component. Data points (in the rest frame) are shown in red; the stellar model denoted with a green dashed line,
the torus with a blue line, and the cold dust component with a dashed light green line.}
\label{fig:cosmosfits}
\end{figure*}

\begin{figure*}
\centerline{
\psfig{file=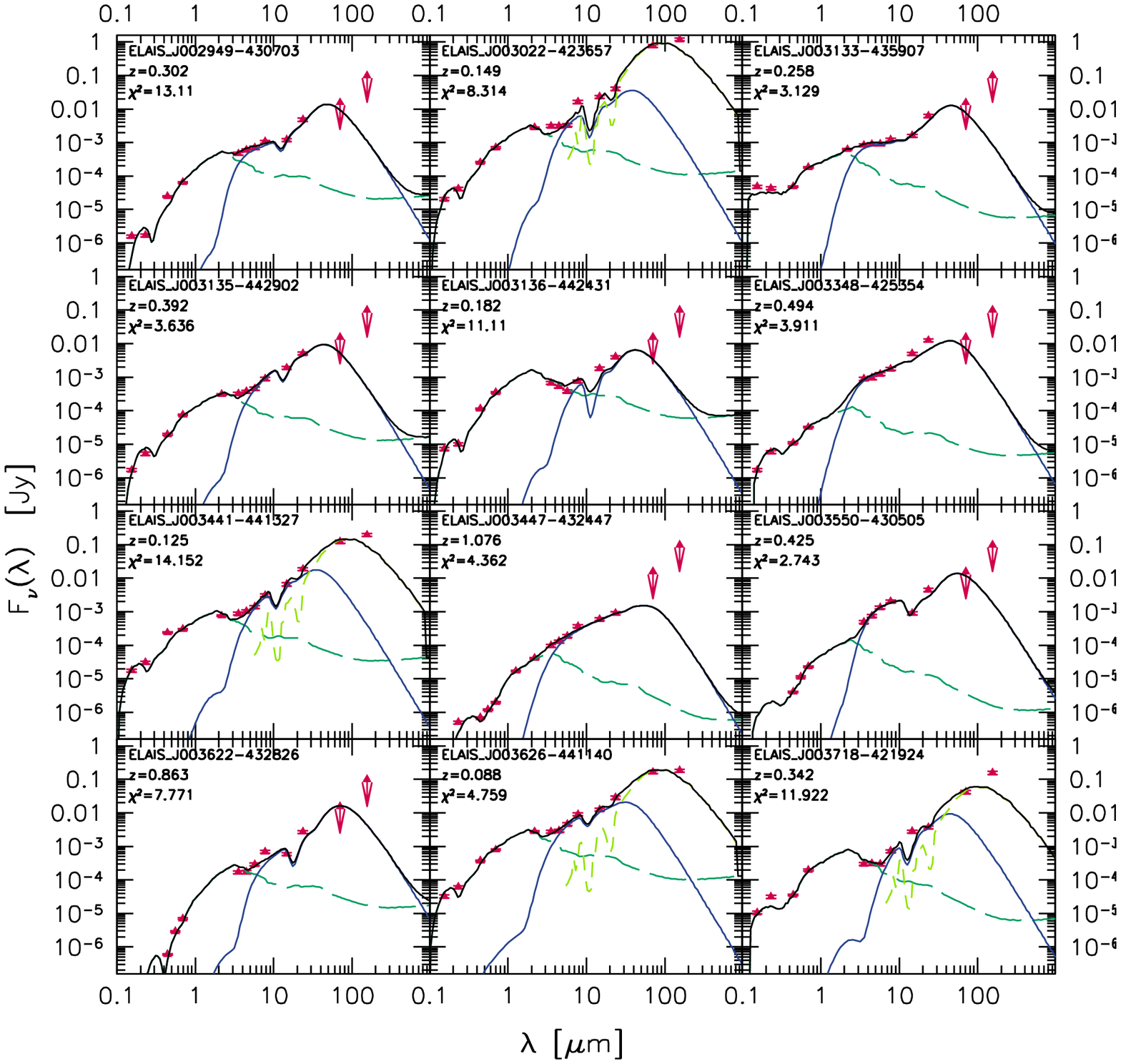,width=.97\textwidth}}
\caption{The best fits for the first 12 ELAIS AGN with reduced $\chi^2<16.0$ that were assigned a
torus component. As in Figs. 5 and 13, data points are shown in red; the stellar model is marked with a green dashed line,
the torus with a blue line, and the cold dust component with a dashed light green line.}
\label{fig:elaisfits}
\end{figure*}

Examples of SED fits of objects assigned an AGN component are shown in
Figs. \ref{fig:cosmosfits} and \ref{fig:elaisfits}, for the COSMOS and
ELAIS samples, respectively. The fits for all objects are available as
online material. 

Note that in the upper and lower left plots in Fig. \ref{fig:cosmosfits}
(objects COSMOS\_J100137.69+002844.1 and COSMOS\_J100249.59+013520.2,
respectively),
the observed 24 \mums points lie well above the model (in both 
cases consisting only of stars and a torus). Such cases indicate the
possible existence of an additional SB component, which we did not use
for these fits because of the lack of observed 70 and/or 160 \mums data
points, as discussed in detail in Section \ref{sec:sedfit}.
Similarly, for the lower right object in Fig. \ref{fig:elaisfits}
(ELAIS\_J003718-421924), the fit fails to reproduce the 160 \mums
point, despite the simultaneous existence of a torus and a SB
components. This case shows that either the SB templates used are not
always adequate or, most probably, that we are lacking a diffuse component
that would account for the cold dust in the host galaxy.

The fitting results show
a clear preference for models with dust density decreasing with the distance from the centre,
with 95\% of the objects finding a better fit with $\beta < 0.0$ models (75\% with $\beta=-0/5$).
A further dependence of the dust spatial distribution on the altitude from the equator was found, 
with 87\% of the objects having a best fit model with $\gamma =6.0$.
The resulting tori sizes do not depend on the type of object (AGN or hybrid) 
and the estimated average inner 
tori radii, $R_{in}$,
are of 0.85$\pm$0.50 
and 0.91$\pm$0.58 pc, for AGN and hybrids, respectively, and with
about 60\% of the objects better matching torus models with
small sizes ($R_{out}/R_{in}$=30).
More than half of the objects were found to have a large covering factor ($>80$\%). A comparative
study on the covering factor derived from the various samples will follow shortly.

\begin{figure}
\centerline{
\psfig{file=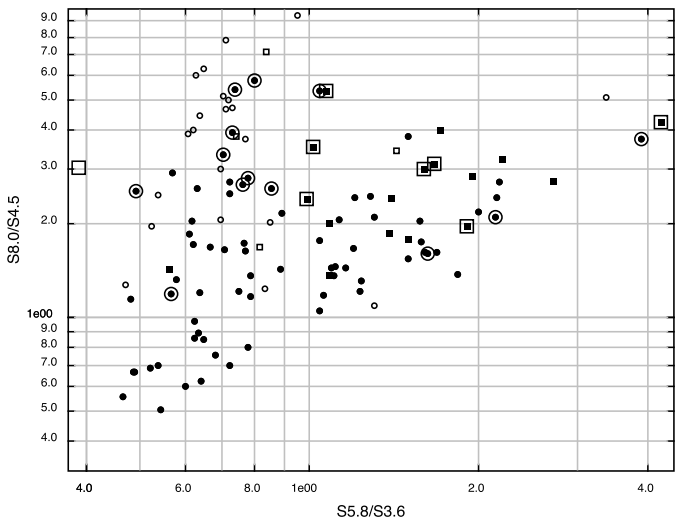, width=10cm}}
\caption{IRAC colour diagram for the COSMOS (circles) and ELAIS (squares) AGN. Small filled symbols denote
an AGN component, large open symbols an SB component, for objects with a good $\chi^2$. Small open 
symbols mark the objects whose fits were not acceptable.}
\label{fig:cosmos_colours}
\end{figure}

Once again, the position of the objects in the IRAC colour diagram (Fig. \ref{fig:colours})
and the contribution of the various components in their global SEDs are in very good agreement.
Fig. \ref{fig:cosmos_colours} shows the IRAC colour diagram for the COSMOS (circles) 
and ELAIS (squares) samples.

The majority of objects with a 70 micron detection and an assigned SB template gather in the
PAH-dominated locus (upper left part of the diagram) while the AGN contribution to the IR
luminosity increases as objects move along the continuum-dominated locus, becoming redder in
the $S_{5.8}/S_{3.6}$ axis.

From the 23 objects composing the MIR selected type 2 ELAIS AGN sample,
21 were found to have a torus component, all of which with $\tau_{9.7} \ge 6.0$,
$\gamma$=6.0 and
$\beta < 0.0$, among which 15 found a better fit with $\beta=-0.5$ templates.
The average estimated sizes of the tori were very similar to those of the COSMOS sample.
The reduced $\chi^2$ distribution for the fits is shown in Fig. \ref{fig:TGhistoChi2}
in dashed lines.

\begin{figure}
\centerline{
\psfig{file=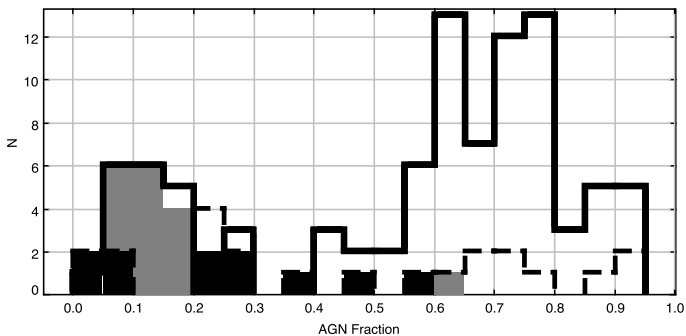,width=10cm}}
\caption{Histogram of the fraction of IR luminosity attributed to the AGN, for objects
from the COSMOS and ELAIS samples..}
\label{fig:cosmoselais_agnFrac}
\end{figure}

The AGN contribution to the IR luminosity for both samples 
is shown in Fig. \ref{fig:cosmoselais_agnFrac}, with the
grayed region representing the objects with an additional 
SB component. The plain (dashed) line and grayed
(black) region correspond to the COSMOS (ELAIS) sample. Again, the existence of
such a component can be, in general, well constrained only in the presence of data points longward $\lambda=24$\mums
and this is valid for only 25\% and 50\% of COSMOS and ELAIS samples, respectively.

\section{COMBINED AGN PROPERTIES}
\label{sec:combi}

We will now attempt to put all the results from the various AGN
samples together and construct a common description of their dust properties.
To this end, we will first present a summary of the
results of Paper 1, that will also be used in this combined study.
The low-$z$ sample will be excluded from this analysis, because of the
unconfirmed (stellar vs. starburst) nature of the objects. 
They will be taken into account, however,
when discussing the star formation, as the results on this issue seem to
be more robust, and only marginally affected by the presence of tori,
as suggested from Fig. \ref{fig:kauff_agnFrac}.

\subsection{Type 1 SWIRE/SDSS quasars}
Paper 1 focused on the study of 278 SDSS/SWIRE quasars and the
properties of dust surrounding them. Comparing the results between
the ``restricted'' run, where only high $\tau_{9.7}$ models were allowed
and the ``full'' run, where optical depths as low as 0.1 were allowed,
the hypothesis of the existence of low optical depth tori could not
be ruled out. In fact, the majority of objects found a better match of 
their SEDs with low $\tau_{9.7}$ models ($\tau_{9.7} < 1)$. The computed
average inner 
radius 
was of $\sim$2.6 pc
with a tendency for higher $R_{out}/R_{in}$i (=100) for the full run
(59\%) with respect to the restricted run (46\%), as reported in Paper 1.
The covering factor depended of course
on the choice of the optical depth but had a relatively flat distribution
in both cases, taking values from as low as $< 0.1$ to as high as 
0.95. As for the dust density, there was no clear preference between
$\gamma=0.0$ and $\gamma=6.0$ models, in any of the runs, while $\beta<0.0$
models were always favoured. The accretion and IR luminosities were
the two better constrained quantities and were almost independent on
the choice of $\tau_{9.7}$. The subsample of 70 objects with 70 \mums
detections enabled an estimation of the contribution of a starburst
to the total IR luminosity, indicating contributions of up to 80\%.

\subsection{Comparative study of type 1 \& 2 AGN}
\label{sec:compare}

Section \ref{sec:dataall} described the observed properties of the combined AGN
sample. Although the sample is not complete,
the combined study of the various AGN sub-samples
composing it, will allow for comparative conclusions. 

\begin{figure*}
\centerline{
\psfig{file=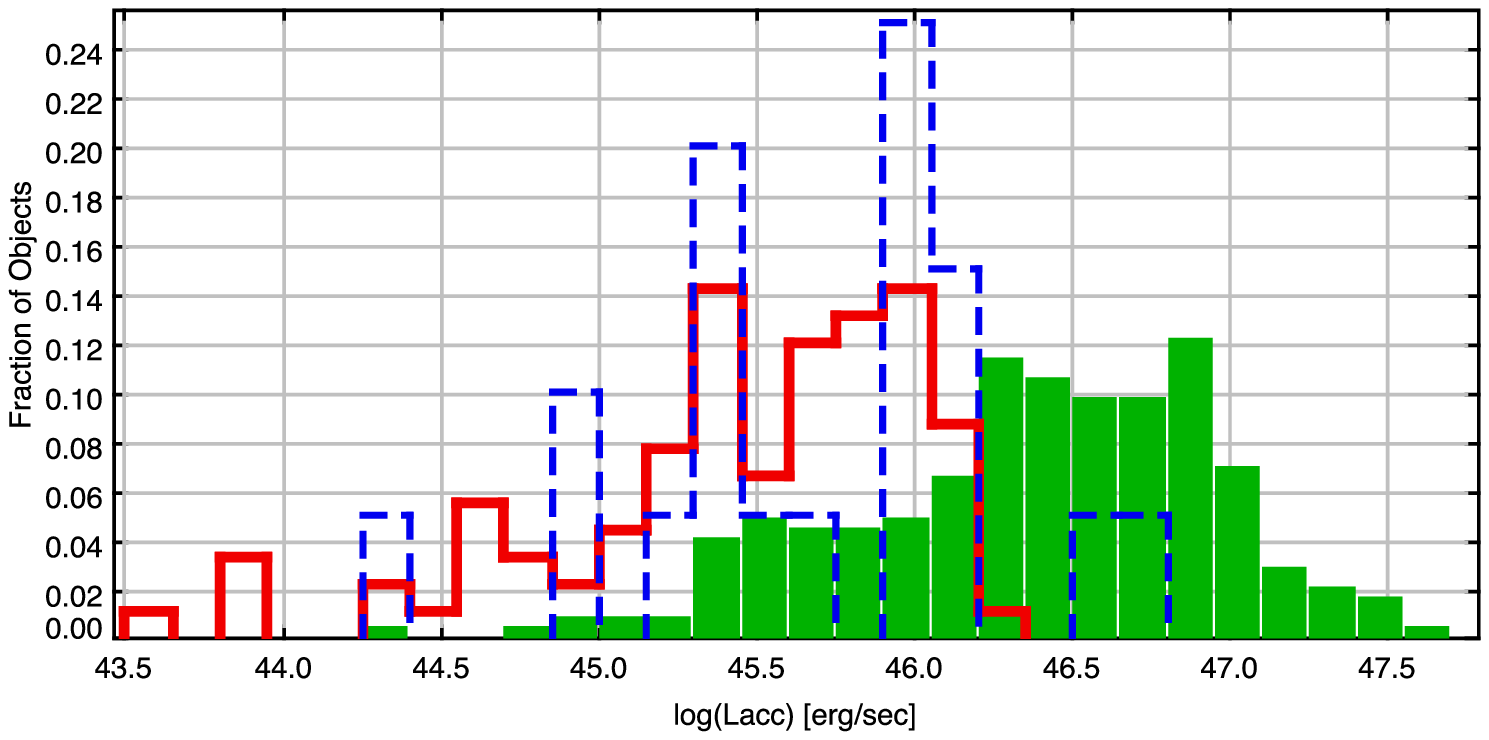,width=10cm}
\psfig{file=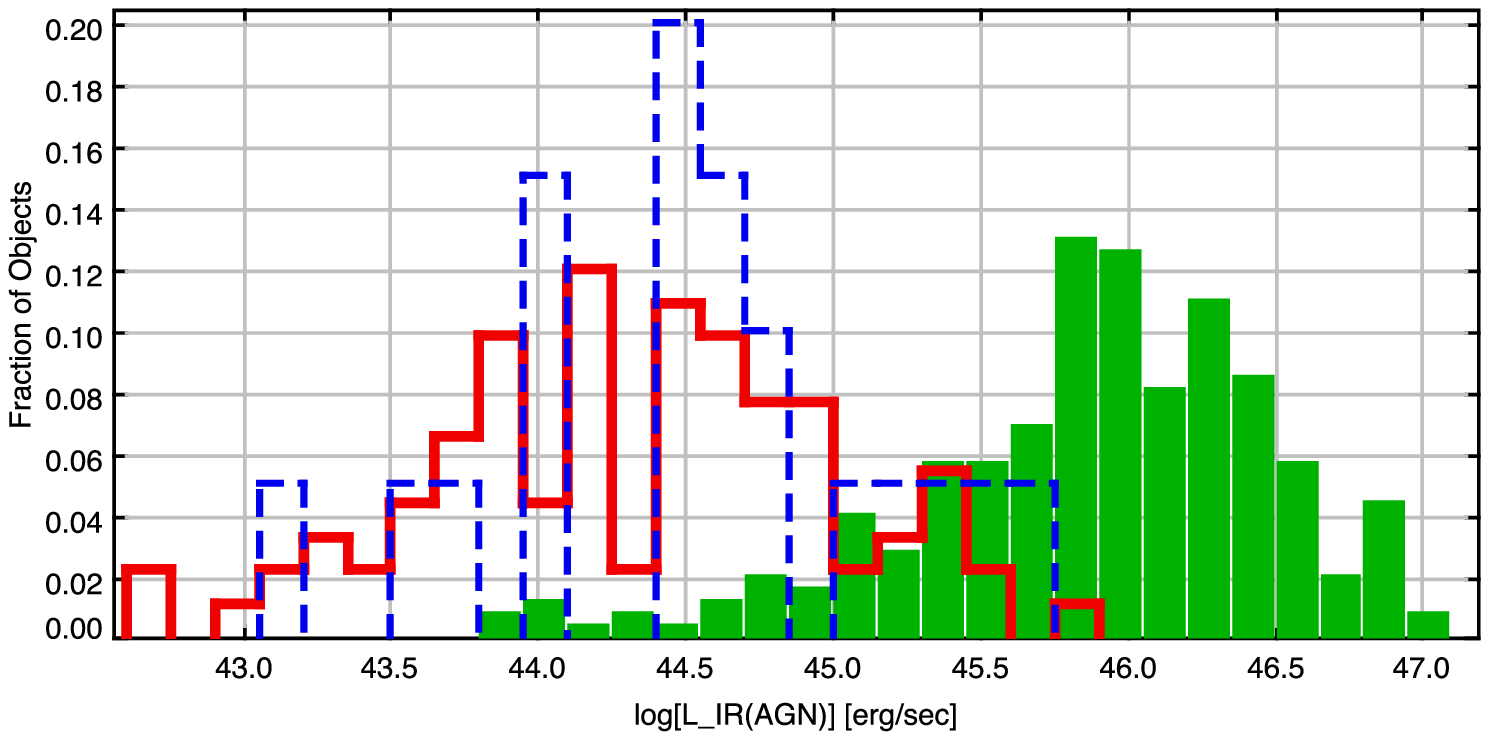,width=10cm}}
\centerline{
\psfig{file=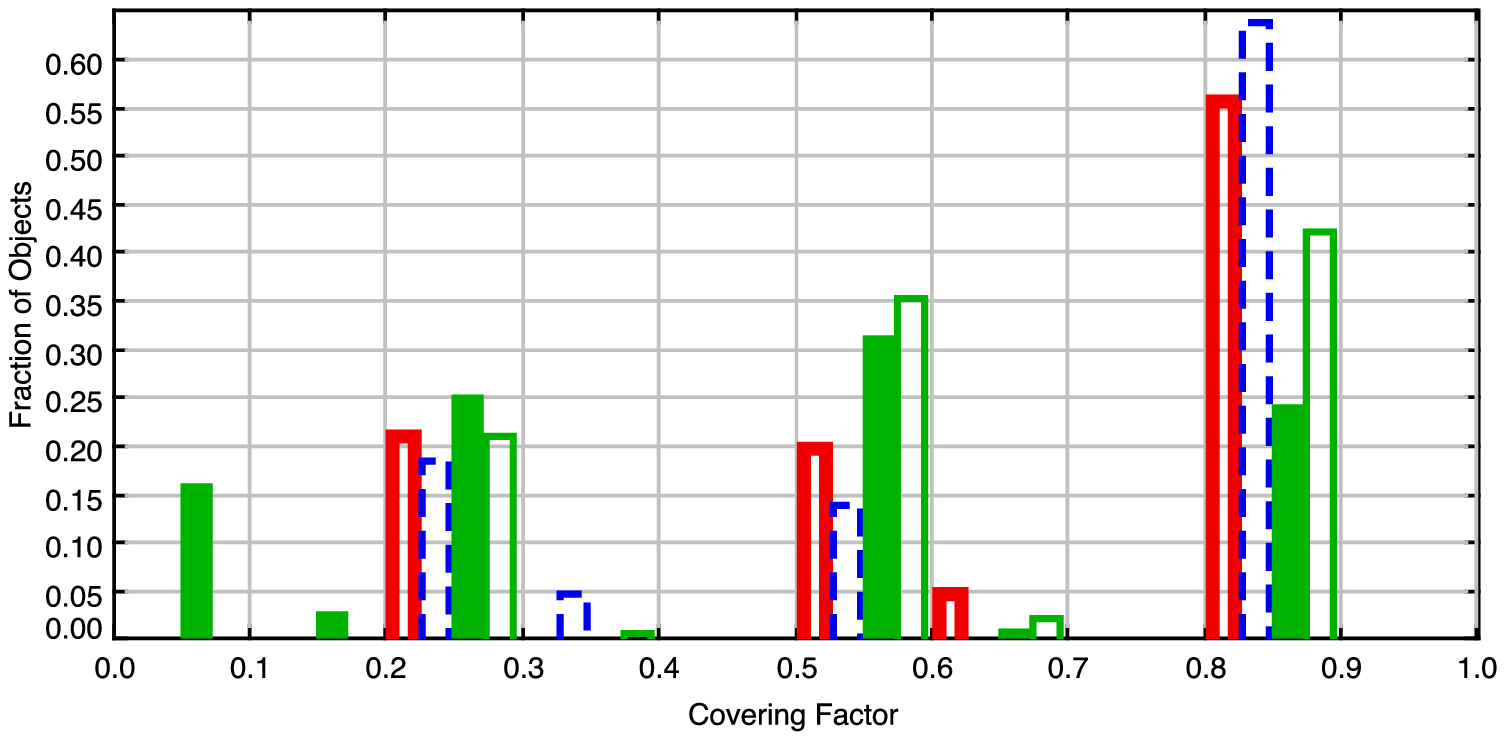,width=10cm}
\psfig{file=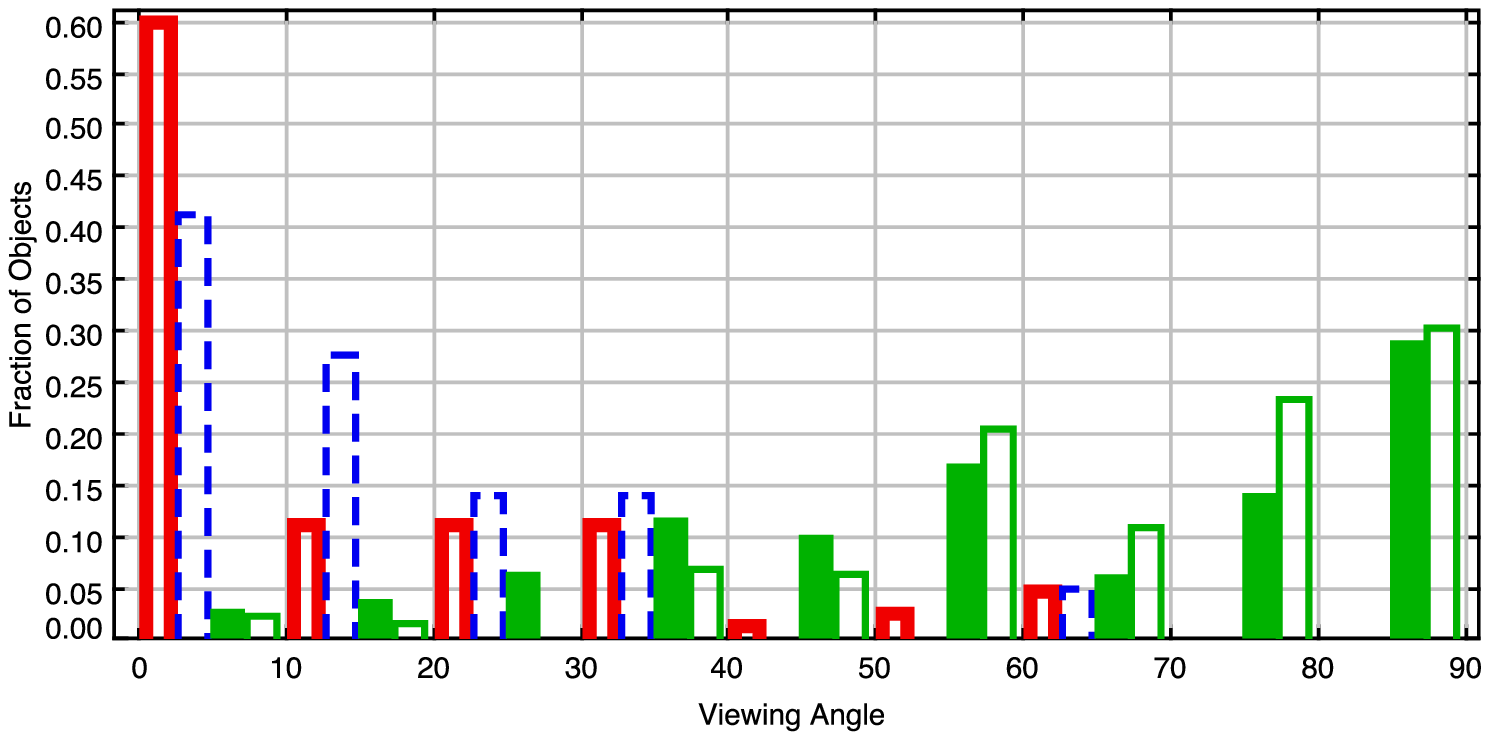,width=10cm}}
\centerline{
\psfig{file=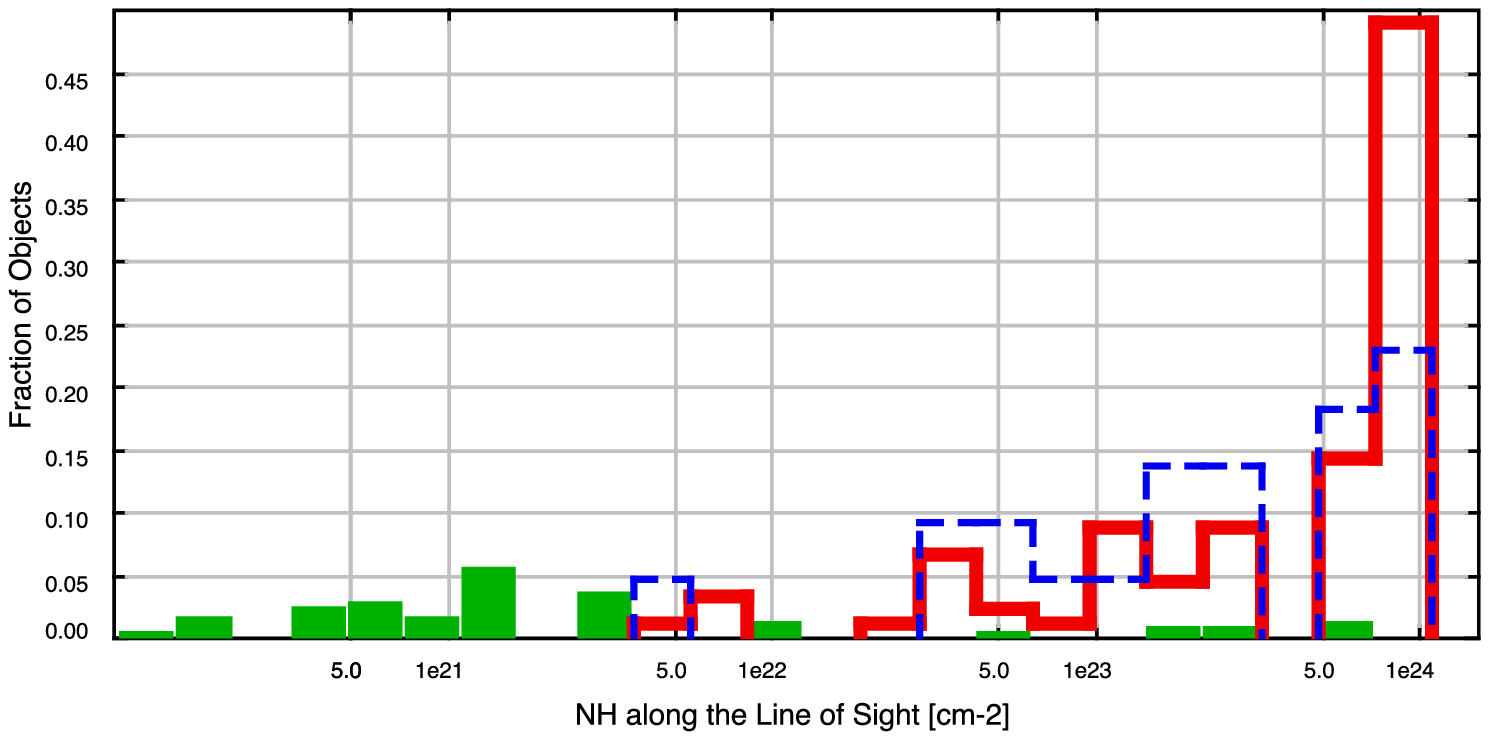,width=10cm}
\psfig{file=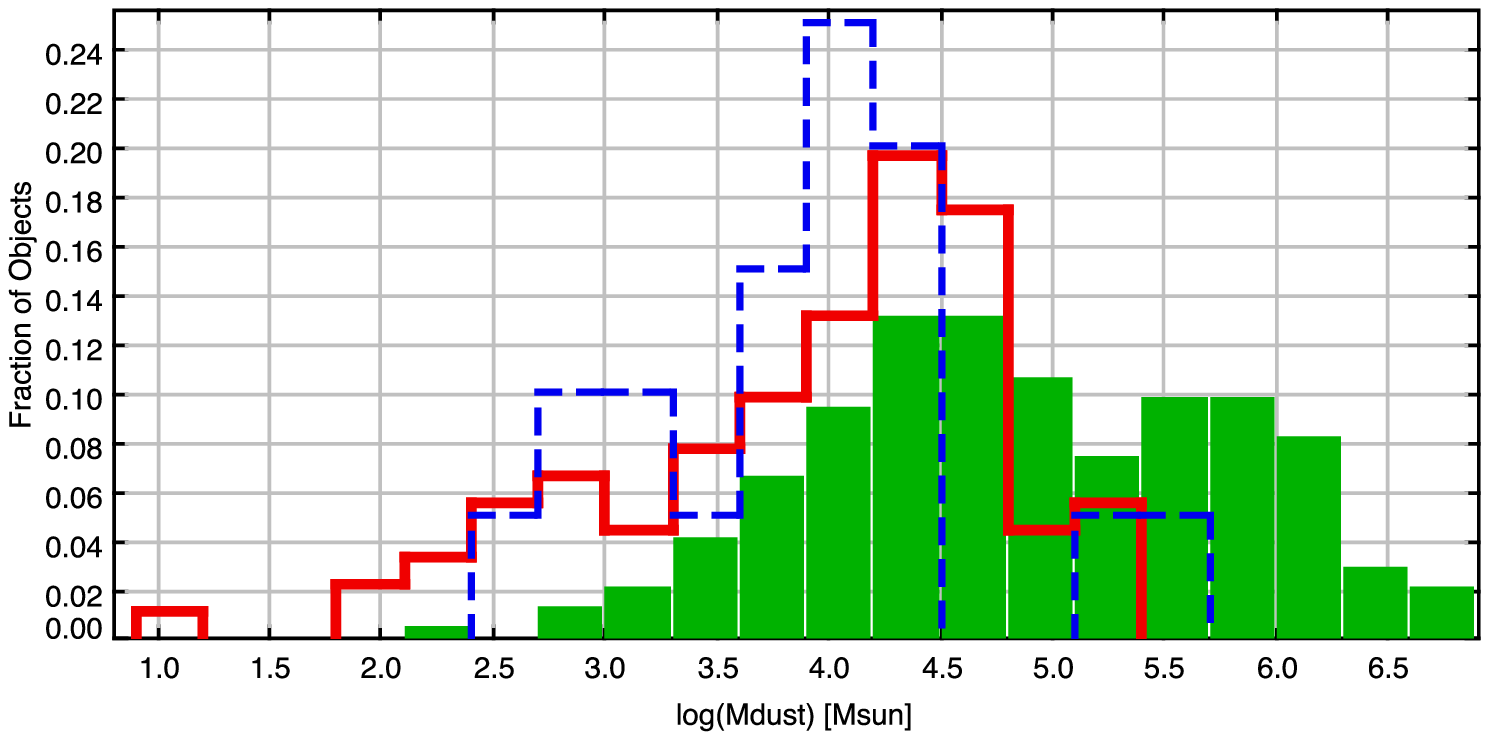,width=10cm}}
\caption{From top to bottom and left to right:
accretion luminosity, $L_{acc}$, and infrared luminosity attributed to
the AGN component,
covering factor, viewing angle, hydrogen column density, and mass of dust,
for the various samples.
Green filled and empty histograms show the distributions for the quasar ``full'' and 
``restricted'' runs, respectively; red open full-line and blue dashed-line histograms
show the results for the COSMOS and ELAIS samples, respectively. The low-$z$ sample
has been excluded.}
\label{fig:compare_props}
\end{figure*}

Fig. \ref{fig:compare_props} shows the distribution of the various
parameters for the different samples. From top to bottom and left to
right, and following the colour-coding introduced in the left panel
of Fig. \ref{fig:colours} we present: the accretion luminosity; and the
IR luminosity attributed to the torus component
(i.e. excluding the stellar emission coming either from the SSPs and/or
the SB components); the covering factor; the viewing angle; 
the hydrogen column density along the line of site; and the mass of
dust confined inside the torus. The open histogram corresponds to the
``restricted'' run, where only high optical depth tori models
($\tau_{9.7} \ge 1.0$) were allowed (Paper 1).

The combined quasar and type 2 AGN samples cover four orders of magnitude in $L_{acc}$,
with the quasar $L_{acc}$ distribution peaking at luminosities of about an order of magnitude
brighter than the type 2s (see Fig. \ref{fig:compare_props} - upper left). 
This observational bias affects, for example, the fraction
of the IR luminosity attributed to the torus component (Fig. \ref{fig:compare_props} - upper
right).

The viewing angle as derived from the fits, shown in the right column, second row, is consistent
with the Unified Scheme postulating that the differences between type 1 and type 2 objects are 
a line-of-sight effect, with the type 2 objects seen through the obscuring material.
Almost all type 2 objects are seen in small viewing angles, i.e. close
to the equator and through large amounts of dust, while type 1 objects are seen in large viewing angles
especially when only tori with high optical depths are considered (open green histogram).
Note that low optical depth tori would result in low obscuration type 2 AGN when seen edge-on.
Fig. 4 in \cite{fritz06}, e.g., shows the emission of an AGN through a $\tau_{9.7}=0.1$ torus
for all lines of sights. While the object is clearly a type 1 when seen face-on, the UV-to-optical
nuclear emission can be absorbed by several orders of magnitude when seen through the dust,
depending on the wavelength and the line of sight.

The majority of type 2 objects are seen through high column densities ($N_H \ge 10^{23} cm^{-2}$),
while the few quasars that are seen through the torus, are actually seen through low optical depth 
and lower than the type 2 column density gas
(Fig. \ref{fig:compare_props}, lower left panel).
Also consistent with the Unified Scheme is the distribution of the mass of dust
(right column, fourth row in Fig. \ref{fig:compare_props}). If the mass of dust can be considered
a linear tracer of gas, this implies that the gas reservoirs in the vicinity of type 1 and type 2
objects are comparable.

Note that the properties of the radio and X-ray (COSMOS) and MIR-selected (ELAIS) AGN are remarkably similar
with the only notable difference being the slightly higher IR luminosities of the torus components
of the ELAIS sample (upper right plot in Fig. \ref{fig:compare_props}), simply reflecting the
selection effects. 

\subsection{The covering factor}

The distribution of the values for the covering factor (seen in the left column, second row
in Fig. \ref{fig:compare_props}) indicates that tori with both high and low covering factors 
are possible in both types of objects. The COSMOS and ELAIS samples show
a stronger tendency for higher values.

The estimated type2:type1 ratios based on the CF distribution of the type2 objects
is $\sim$2-2.5:1; the same ratio is found when considering the restricted run of the
quasar sample. The full type 1 run, however, suggests a ratio of 1:1, with a mean value
of the covering factor of $\sim$0.47, very close to the value of 0.4 computed by
\cite{rrobinson08} for the entire SWIRE quasar population. This
low value, seemingly in conflict with the Unified Scheme, actually reflects the
difference in the accretion luminosity between the type 1 and type 2 samples (Fig.
\ref{fig:compare_props}), already discussed in Section \ref{sec:compare}. 

\begin{figure*}
\centerline{
\psfig{file=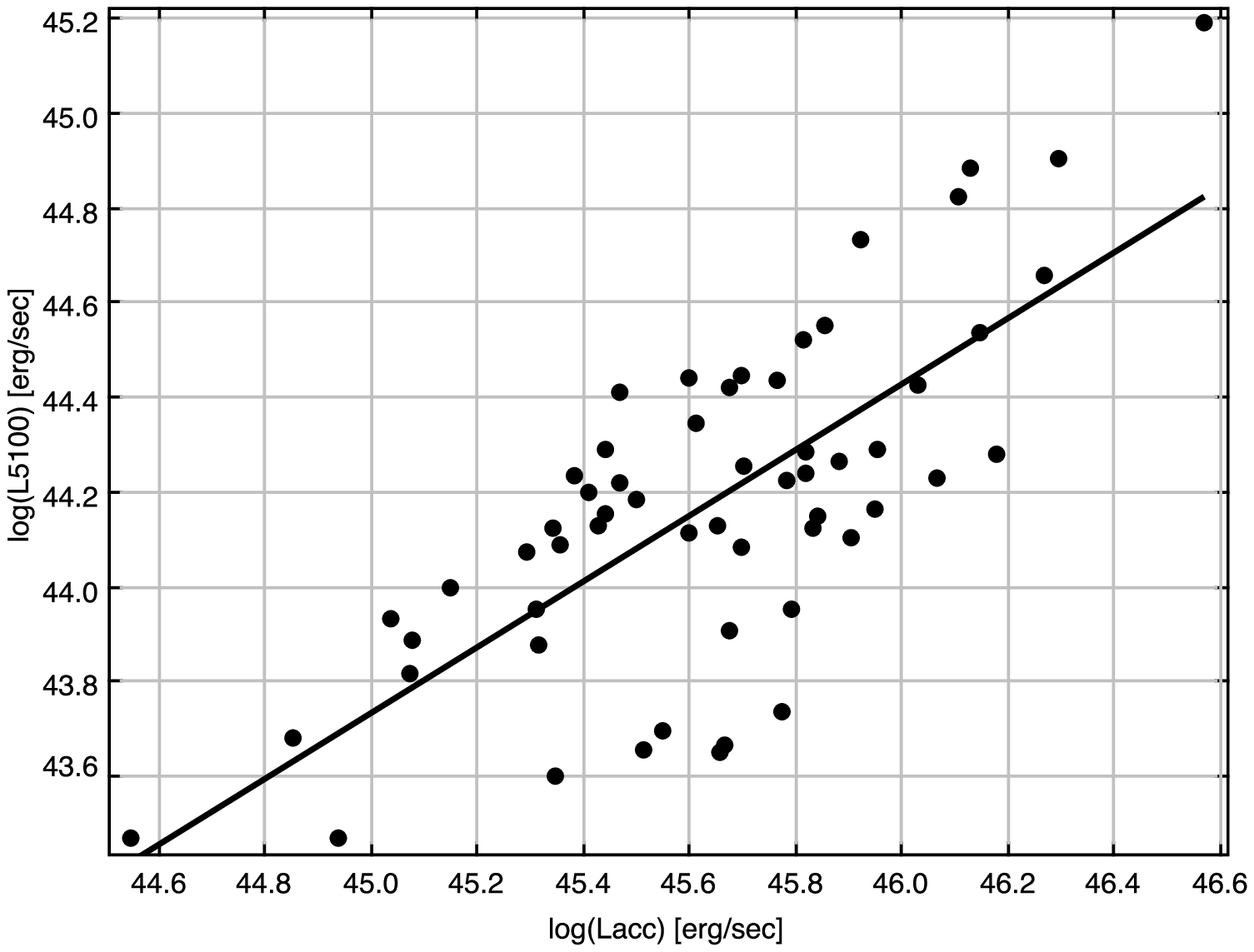,width=10cm}
\psfig{file=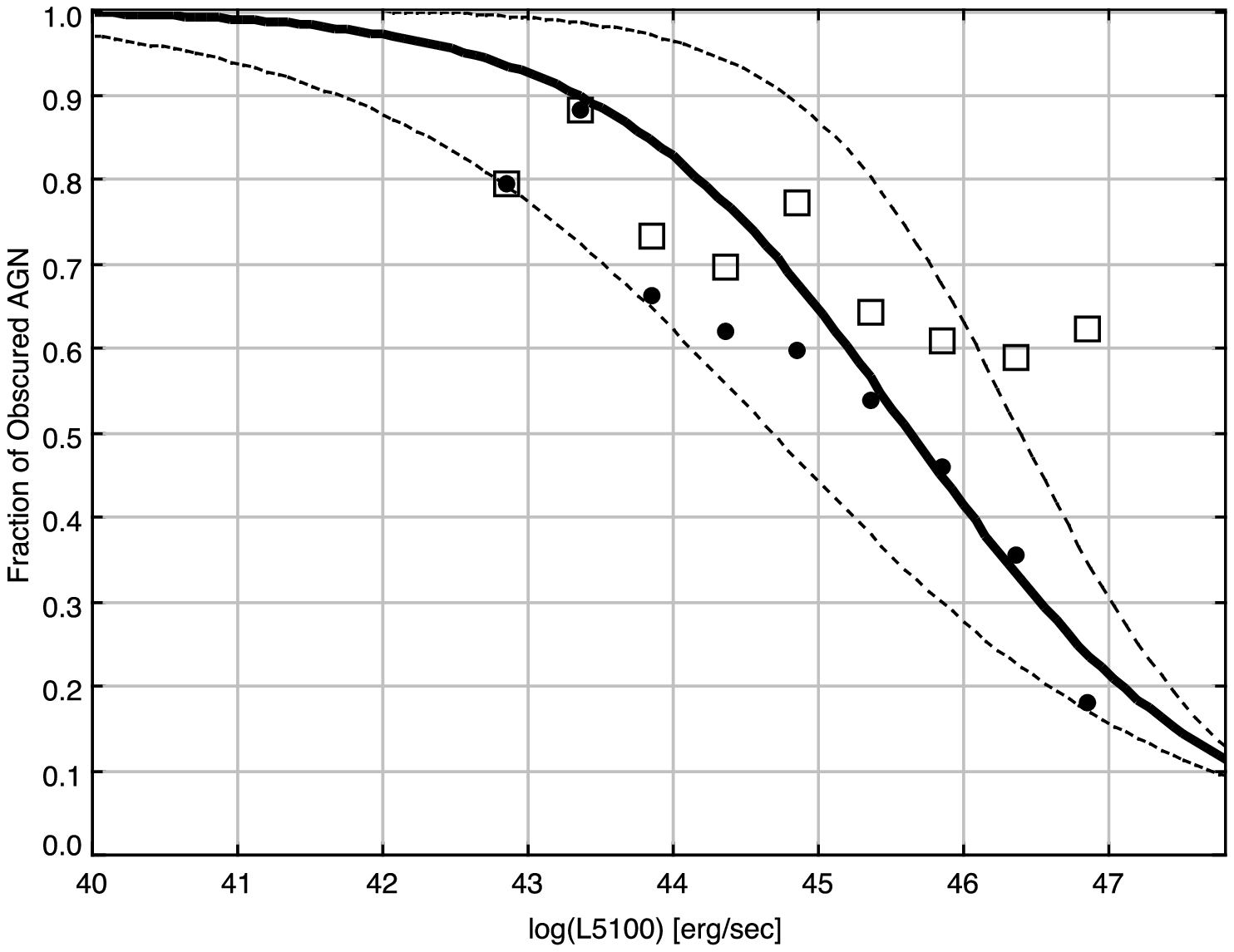,width=10cm}}
\caption{Left panel: $L_{5100}$ measured on the quasar SDSS spectra for objects with redshift lower than
$\sim$0.8, versus $L_{acc}$, computed from the SED fitting. The linear correlation, shown in 
a solid line, is given by $log(L_{5100})=0.692 \times log(L_{acc}) - 1.5767$. Right panel: fraction
of obscured AGN in bins of $L_{5100}$. Filled circles (open squares) represent the results of the full 
(restricted) run. The solid line represents the fraction of obscured AGN as found by Maiolino et al. (2007), 
while the dashed lines trace the uncertainties due to bolometric corrections.}
\label{fig:fracobscagn}
\end{figure*}

From the type 1 quasars with redshifts below $\sim 0.8$ we can measure the flux at
5100 \AA, which we can then scale with the accretion luminosity, $L_{acc}$, as seen in
the left panel of Fig. \ref{fig:fracobscagn}. Assuming this relation holds for all redshifts
as well as for type 2 objects, we can translate $L_{acc}$ to $L_{5100}$. We can then compute
the fraction of obscured AGN in bins of $L_{5100}$, shown in the right panel of Fig.
\ref{fig:fracobscagn}. The black circles and open squares make use of the results
of the full and restricted runs, respectively. The full and two dashed lines show the results and uncertainties
(due to bolometric corrections) obtained by \cite{maiolino07} who conducted a study of 25 high luminosity high redshift ($2 < z < 3.5$) 
quasars with IRS spectroscopy.
The observed trend, i.e. the decreasing average torus covering factor with increasing 
AGN luminosity, is in support of the receding torus paradigm, suggested by \citealt{lawrence91}.

Despite the crude conversion of $L_{acc}$ to $L_{5100}$ our results making use of the full run
are in very good agreement
(i.e. within the uncertainties) with those derived by \cite{maiolino07} based on totally
independent method and sample, while the results based on the restricted run, even though still
consistent (excluding the brightest two bins) show a quite different trend.
We can therefore still not conclude on the relative occurence of low $\tau_{9.7}$ tori
with respect to high $\tau_{9.7}$ tori but, based on the above, 
we can still provide some evidence for their existence.

\begin{figure}
\centerline{
\psfig{file=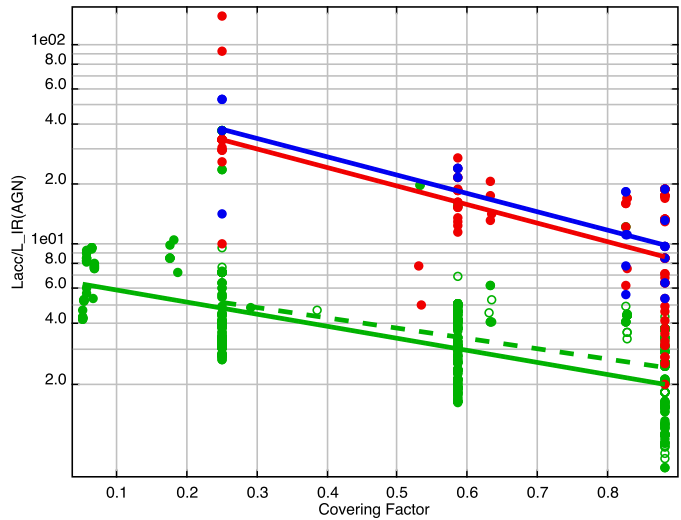,width=10cm}}
\caption{Accretion-to-IR luminosity attributed to the AGN component as a function of
the covering factor. The green solid and dashed lines 
obtained as linear fits corresponding to the full and restricted runs, respectively, for type 1 objects,
while blue and red lines correspond to ELAIS and COSMOS type 2 AGN, respectively.}
\label{fig:lacclircf}
\end{figure}

\begin{figure}
\centerline{
\psfig{file=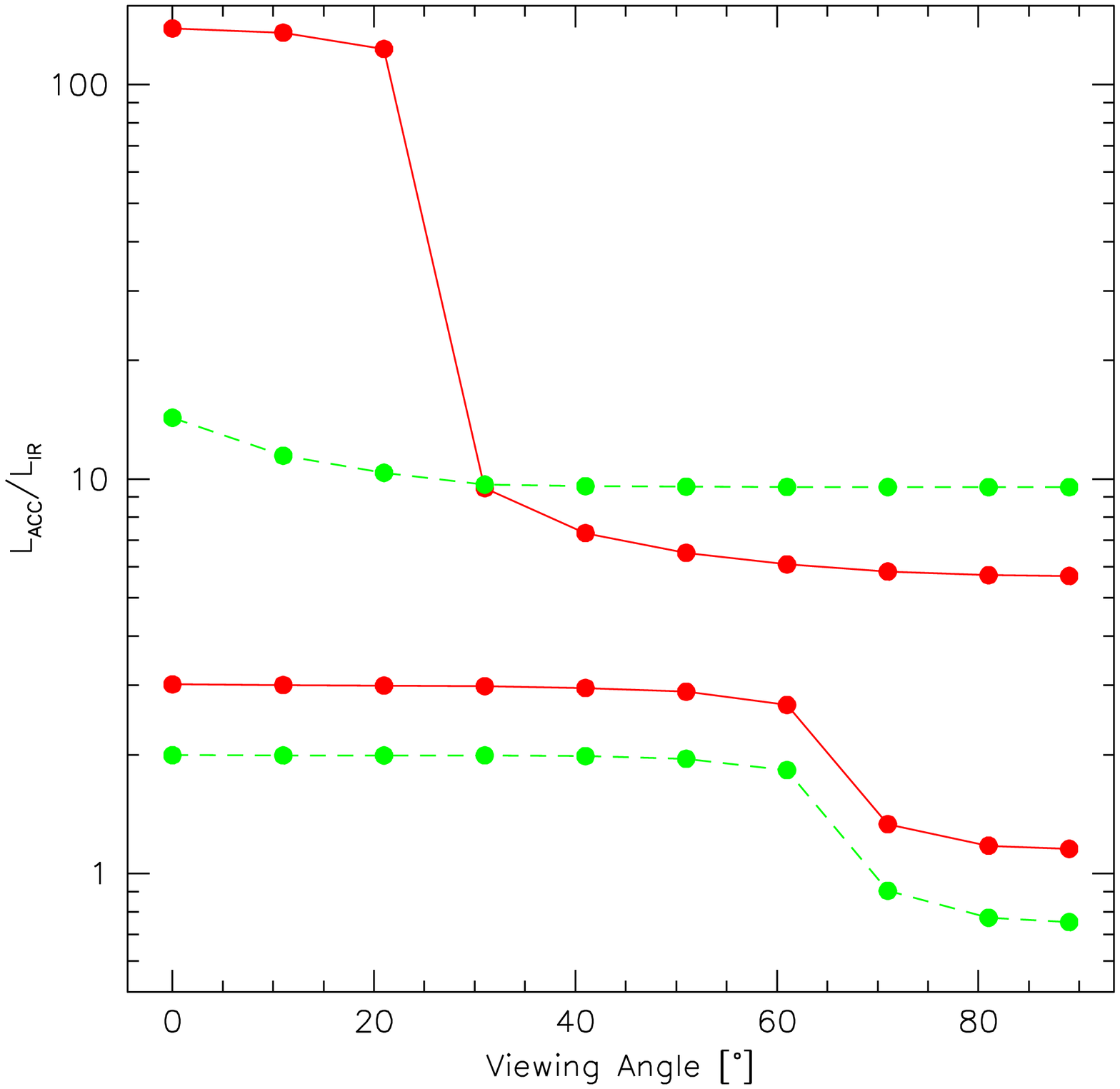,width=10cm,height=8cm}}
\caption{Accretion-to-IR luminosity attributed to the torus as a function of
viewing angle for the type 1 (green dashed lines) and 2 (red solid lines) objects with
the highest and lowest $L_{acc}/L_{IR}(AGN)$ in Fig. 19. The differences in this fraction 
are mainly attributable to dust self absorption which is more efficient for type 2 line of sights.}
\label{fig:lacclirvang}
\end{figure}

Previous studies (e.g. \citealt{maiolino07}) assume that the ratio between the primary 
AGN radiation (here measured by $L_{acc}$) and the thermal infrared emission attributed
to the AGN is a direct indicator of the torus covering factor.
Our findings confirm this assumption, as shown in Fig. \ref{fig:lacclircf}
where the accretion-to-IR AGN luminosity of each individual object of each sample
is plotted against the computed covering factor. The slope of the correlation
is very similar for type 1 and type 2 samples. Type 2 objects, however, have 
considerably larger $L_{acc}/L_{IR}(AGN)$ ratios than the quasars. This ``jump'' in the 
values is due to the different viewing angles and occurs right
when the line of sight intercepts the obscuring torus.
In order to illustrate this effect, we show the $L_{acc}/L_{IR}(AGN)$ ratios
as a function of the viewing angle for the extreme models of both type 1 and 2
objects (higher and lower red points; higher and lower green points in Fig.
\ref{fig:lacclircf}) in Fig. \ref{fig:lacclirvang}.

\subsection{Starburst activity in AGN}
\label{sec:sfr}

The study of objects with at least one data point at wavelengths
longward of $\lambda=24$ \mums shows that the IR emission
of AGN can not be attributed to the torus component alone. In the vast
majority of cases, and in order to reproduce the 70 (and, whenever available, 
160) \mums points, an additional SB component
was necessary. Furthermore, in all those cases the contribution of the AGN
emission in the infrared is typically
smaller than 50\%, as seen in Figs. \ref{fig:kauff_agnFrac} and 
\ref{fig:cosmoselais_agnFrac} for the type 2 objects, and a bit higher
in type 1 quasars (Paper 1, Fig. 12).

\begin{figure}
\centerline{
\psfig{file=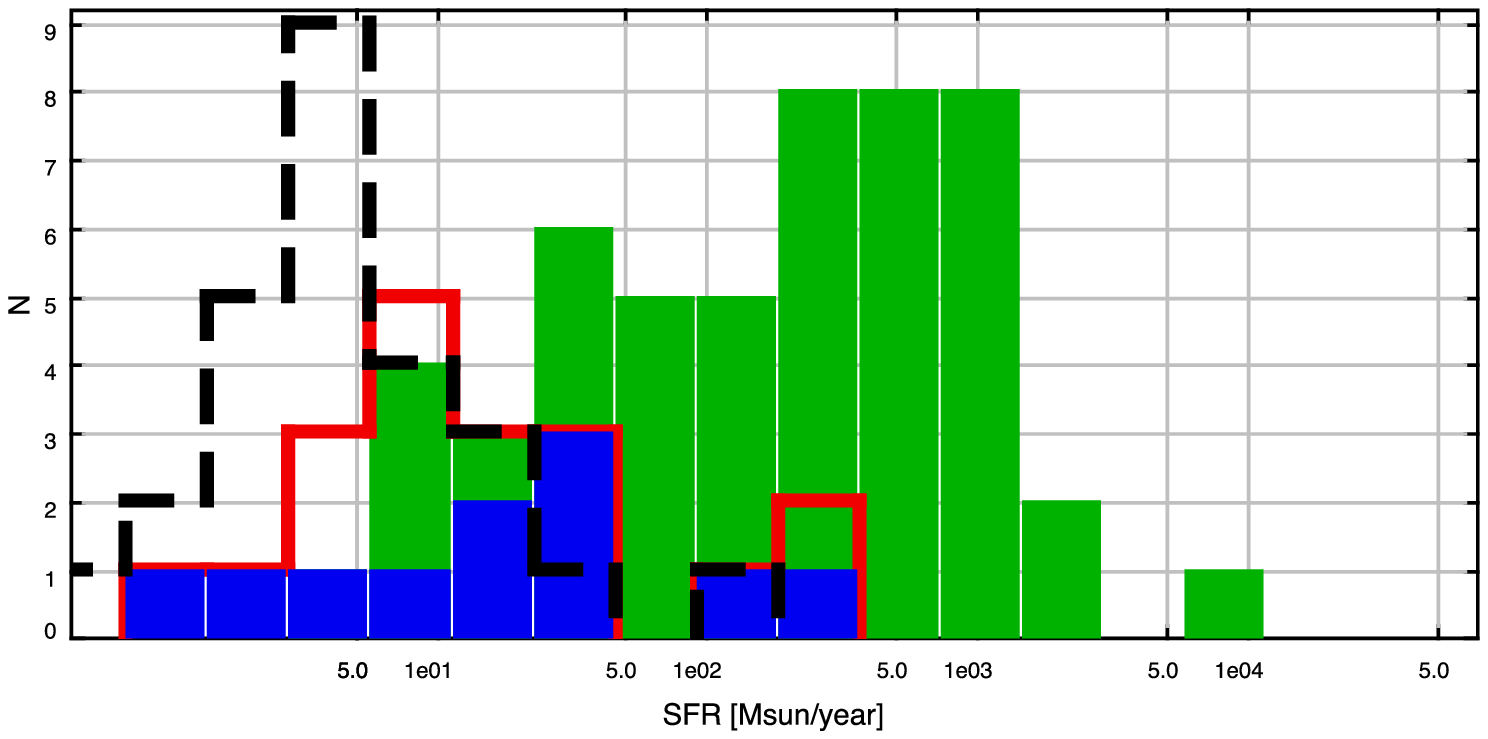,width=10cm}}
\caption{SFRs estimated from the Kennicutt (1998) formalism for the various samples.}
\label{fig:sfr}
\end{figure}

Fig. \ref{fig:sfr} shows the estimated SFRs for the various samples (with the same 
colour coding as in Fig. \ref{fig:colours}) computed from Eq. \ref{eqn:sfr}
for all objects with an SB and an AGN components.
The average SFR for the low redshift sample is about 10 $M_{\odot}$/yr, while those computed
for the COSMOS and ELAIS samples are of $\sim$40 $M_{\odot}$/yr but with a larger span.
That of type 1 quasars (shown in green), however, has an average SFR of 115 $M_{\odot}$/yr, with some cases 
reaching SFRs of the order of 10$^3$ $M_{\odot}$/yr.
These values are well in agreement with the findings of \cite{serjeant09}, a study based on stacking 
analysis of a variety of quasar samples.
These values do not necessarily imply that star formation activity is stronger in type 1 quasars {\it in general}.
More likely, it reflects the observational biases of the samples, both in terms of average
redshifts and luminosities (Figs. \ref{fig:zhisto2} and \ref{fig:compare_props}).

\begin{figure}
\centerline{
\psfig{file=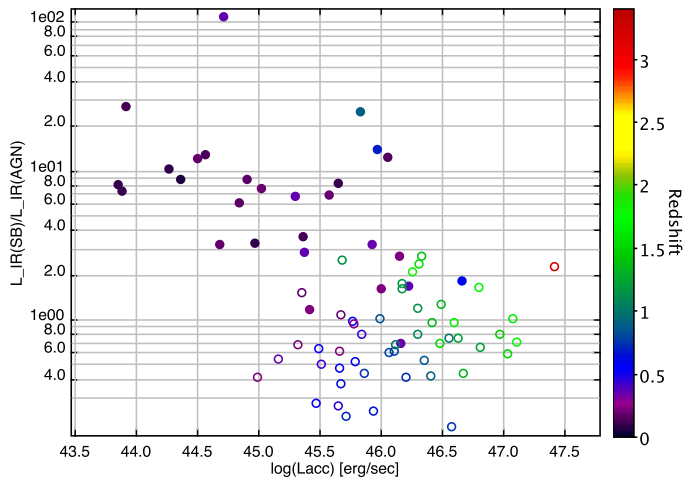,width=10cm}}
\caption{Starburst-to-AGN infrared luminosity ratio as a function of the accretion luminosity, $L_{acc}$,
and the redshift. Filled and open symbols depict the type 2 and type 1 objects, respectively.}
\label{fig:LsbLagnLaccZ}
\end{figure}

The starburst-to-AGN luminosity ratio for the combined sample shows a slight tendency to
decrease with increasing accretion luminosity, as seen in Fig. \ref{fig:LsbLagnLaccZ}.
This is in apparent agreement with previous works on type 1 quasars alone
(again by \citealt{maiolino07}), and if confirmed for the entire sample, would imply
that type 1 and type 2 objects behave in the same way in this respect.  On the other hand,
we can not assert beyond doubt that there there is no implicit dependency on the redshift.
In fact, examining the quasar sample alone (open symbols) we would be tempted to claim
an {\it increase} with $L_{acc}$, but which is likely attributed to the $L$ - $z$
degeneracy.

\section{DISCUSSION}
\label{sec:discuss}

This work focuses on the IR properties type-2 AGN, the type 1 issues having already 
been discussed in a previous paper \citep{hatzimi08}. The collection of data
used for this paper comes from: 1) a low-redshift ($0.02<z<0.2$) AGN sample, 
selected via standard emission-lines ratios criteria; 2) a higher redshift ($0.1<z<1.4$) 
sample from the COSMOS and ELAIS fields, selected with X-ray and Mid-Infrared
criteria.
The observed SEDs were reproduced by means of a three component model, including
emission from stars, hot dust from AGN and warm dust from starburst activity, and
acceptable fits were obtained for the majority of the objects.

Despite the fact that the samples we analyzed were not chosen homogeneously, a comparison 
of the behaviour between the physical properties of the type 1 and type 2 AGN, shows an overall 
agreement. In Paper I we found the covering factor to have a broad trend towards lower values 
as the accretion luminosity, i.e. the power of the central engine, increased with 
small, on-average, covering factors. 
Applying the $L_{acc}$ versus $L_{5100}$ correlation derived for quasars in Paper I, we derived the 
flux at 5100 \AA \ for the type 2 samples and checked
the correlation between the optical luminosity and the covering factor, which in turn, can 
be converted into a type1-to-type2 fraction. The results provided by the SED fitting analysis, 
show a very good agreement with the relation presented in Maiolino et al. (2007), who find a correlation 
between the fraction of obscured AGN as a function of the optical luminosity (i.e. at 5100 \AA). Another 
indication of the dusty torus was thought to be ratio between the accretion luminosity and the torus 
infrared luminosity. In fact, the latter is just reprocessed radiation which scales linearly with the 
primary source: what makes the difference is the fraction of ``heating radiation'' which is intercepted 
by the dust, which again depends on the covering factor. We find a very 
similar trend for these quantities, in both type 1 and type 2 objects, also showing how differences 
in the amount of obscuration are very well explained by {\it dust self-absorption}, i.e. thermal dust emission
absorbed by dust iteself.

For the low-$z$ sample, we found no evidence of emission from an AGN 
component for $\sim 70$\% of the objects. Although we cannot rule out the absence of
an AGN from these sources, we can set an upper limit to its luminosity since it 
is not observed at mid-infrared wavelengths where its emission is the strongest as compared
to both that of the stellar and of the starburst-heated dust. This sample suffers
from high contamination from the stellar emission component, even at mid-infrared wavelengths
where (on the contrary) it's the AGN component that usually dominates when present. 
The question to ask, therefore, is what can be really constrained with three or four data points. 
One of the most reliable quantities should be the optical 
depth, since low values would make the torus emission stand over the stellar (SB or stars) 
in the MIR. In this respect, the low-$z$ sample may be too 
generous in selecting really active AGN. There is, sometimes, no evidence for a MIR excess 
at all with respect, for example, not only to a ``normal'' starburst emission, but also to 
a passively evolving galaxy. The absence of any evidence for 
a hot dust emission in the MIR, which is the place where the stellar component is less important 
and the AGN contribution is increasingly brighter, will remove the AGN component from the fits even though
this could also be turned into an upper limit. In the cases where there is a MIR excess with respect 
to a pure stellar emission, we explore two possibilities: torus and PAH emission. In some cases
a better fit was obtained adding PAH -i.e. starburst component alone- to the model 
instead of AGN which should instead dilute the PAH emission \citep{lutz98}.

Comparison between clumpy and smooth tori models (e.g. \citealt{dullemond05})
indicate that globally the SEDs produced by the two models are quite similar,
but with some details characteristic for one or the other model:
the silicate feature observed in absorption in objects seen edge-on is shallower
for clumpy models, the average near-IR flux is weaker in smooth models and
the clumpy models tend to produce slightly wider SEDs at certain inclinations.
Furthermore, clumpy models can produce very small tori sizes with
$R_{out}/R_{in} \sim 5-10$ \citep{nenkova08}, while still producing a broad MIR
emission. Subsequently, the selection of a smooth torus for the present study
might have resulted in overestimated tori sizes without, however, jeopardising
the estimates on the properties related to the IR emission or our conclusions
on the Unification Scheme.
SED fitting may not be a sensitive enough method to distinguish
between the differences introduced by the two approaches, because of 
the width of the filters and the scarce sampling of the observed SEDs, but also 
because the various characteristics of the torus component can be altered or 
diluted by, for example, the presence of a starburst component. Notwithstanding 
these limitations, SED fitting is still the best tool available for extracting 
the maximum information from large photometric AGN samples and is now proving 
to be a powerful technique in relating the dust properties to the accretion 
properties as well as the properties of the larger host galaxy.

Because AGN are of order hundred times less numerous than galaxies and also
prone to selection biases, such AGN studies are only possible with the advent of
multi-wavelength surveys that probe large volumes and allow the construction
of well-samples SEDs. Even though we cannot address all issues related to dusty tori, 
our understanding of their properties
has been greatly improved over the last decade thanks to the Spitzer Observatory.
This space infrared telescope allowed, among other things, the construction of SEDs 
of hundreds of AGN of all types
thanks to the IRAC and MIPS photometry (e.g. \citealt{franceschini05}; 
\citealt{hatzimi08}; \citealt{richards06}; \citealt{polletta08}). Spitzer also
opened the doors to detailed and coherent studies of the interplay between AGN and starburst activity
(e.g. \citealt{hernan09} and references therein), paving the way for Herschel.

In fact, Herschel with its two cameras/medium
resolution spectrographs (PACS; \citealt{poglitsch09} and SPIRE; \citealt{griffin09}) 
and a very high resolution heterodyne spectrometer (HIFI; \citealt{degraauw09}) will 
be the first space facility to completely cover
the range between 60 and 670, where the the bulk of energy is emitted in the Universe,
allowing for a more detailed sampling of the observed FIR SEDs and the cold dust emission.
Large parts of the Herschel key science will focus on the formation and evolution of 
galaxies and the studies of star formation.
HerMES, the Herschel Multi-tiered extragalactic Survey,
is the largest project that will be conducted by Herschel, with dedicated $\sim$900
hours that will map over 70 square degrees including most of the SWIRE fields 
(where most of the IR photometry of all the objects in this study comes from) as well
as other extragalactic survey fields covered by several missions in all wavelengths
(e.g. COSMOS or the Groth Strip), and will address among others the issue of AGN
and starburst connection (\citealt{griffin06}; \citealt{hatzimi07}).
Additionally with Herschel, the Astrophysical Terahertz Large Area Survey (ATLAS) will
cover about 500 square degrees and will observe all SDSS quasars with redshift
$z < 0.2$ as well as {\it all} the brightest FIR SDSS quasars (with luminosities
10 times larger than the mean), amounting to $\sim$330 individual detections
of quasars in the area covered \citep{serjeant09} (a factor of $\sim$5 more than
all the SDSS quasars with 70 \mums detections).
These studies will allow us to significantly improve our understanding of the AGN phenomenon,
disentangle the contribution of cold and hot dust emission in their SEDs and study
the concomitant occurence of nuclear activity and star formation.

\noindent
\vspace{0.75cm} \par\noindent
{\bf ACKNOWLEDGMENTS} \par

\noindent This work is based on observations made with the {\it Spitzer Space Telescope},
which is operated by the Jet Propulsion Laboratory, California Institute of
Technology under NASA contract 1407.
Support for this work, part of the Spitzer Space Telescope Legacy Science
Program, was provided by NASA through an award issued by the Jet Propulsion
Laboratory, California Institute of Technology under NASA contract 1407.

Funding for the creation and distribution of the SDSS Archive has been provided
by the Alfred P. Sloan Foundation, the Participating Institutions, the National
Aeronautics and Space Administration, the National Science Foundation, the U.S.
Department of Energy, the Japanese Monbukagakusho, and the Max Planck Society.
The SDSS Web site is http://www.sdss.org/.

This publication makes use of data products from the Two Micron All Sky Survey, 
which is a joint project of the University of Massachusetts and the Infrared 
Processing and Analysis Center/California Institute of Technology, funded by the 
National Aeronautics and Space Administration and the National Science Foundation.

This work made use of Virtual Observatory tools and services, namely TOPCAT
(http://www.star.bris.ac.uk/~mbt/topcat/) and VizieR (http://vizier.u-strasbg.fr/cgi-bin/VizieR).

We would like to thank R. Maiolino for kindly providing material for Fig. \ref{fig:fracobscagn}.

We would also like to thank the anonymous referee for their in depth study of the paper
and the subsequent comments that, we believe, greatly improved the manuscript.

\end{document}